\documentclass[eqsecnum,twocolumn,tightenlines,floats,floatfix,prc,nofootinbib]{revtex4}

\def\bea{\begin{eqnarray}}
\def\eea{\end{eqnarray}}

\def\be{\begin{equation}}
\def\ee{\end{equation}}
\def\ba{\begin{eqnarray}}
\def\ea{\end{eqnarray}}

\usepackage{graphics}
\usepackage{graphicx}
\usepackage{epsf} 
\usepackage{amsmath}
\usepackage{amssymb}
\usepackage{slashed}

\def\sfrac#1#2{{\textstyle \frac{#1}{#2}}}

\begin{document}

\phantom{0}
\vspace{-0.2in}
\hspace{5.5in}
\parbox{1.5in}{ } 

\vspace{-1in}

\title
{\bf Timelike $\gamma^\ast N \to \Delta$ form factors and $\Delta$ Dalitz decay}
\author{G.~Ramalho$^{1}$  and M.~T.~Pe\~na$^{1,2}$
\vspace{-0.1in}  }

\affiliation{
$^1$CFTP, Instituto Superior T\'ecnico,
Universidade T\'ecnica de Lisboa,
Av.~Rovisco Pais, 1049-001 Lisboa, Portugal
\vspace{-0.15in}}
\affiliation{$^2$Physics Department, 
Instituto Superior T\'ecnico,
Universidade T\'ecnica de Lisboa,
Av.~Rovisco Pais, 1049-001 Lisboa, Portugal}

\vspace{0.2in}
\date{\today}

\phantom{0}

\begin{abstract}
We extend a covariant model,  tested before
in the spacelike region 
for the physical and lattice QCD regimes, to a calculation of
the $\gamma^\ast N \to \Delta$ reaction in the timelike region,
where the square of the transfered momentum, $q^2$, is positive 
($q^2>0$). 
We estimate 
the Dalitz decay $\Delta \to N e^+ e^-$
and the $\Delta$ distribution mass distribution function. 
The results presented here can be used to simulate 
the $NN \to NN e^+ e^-$
reactions at moderate beam kinetic energies. 

\end{abstract}

\vspace*{0.9in}  
\maketitle

\section{Introduction}

Electromagnetic reactions which induce excited states of the nucleon 
are important tools to study hadron structure, and define 
an intense activity at modern accelerator facilities, 
namely at MAMI, MIT/Bates and Jefferson Lab. 
An enormous progress 
in these experimental studies has been achieved in the recent years, 
leading to very accurate sets of data on 
$\gamma^\ast N \to N^\ast$ excitation reactions, for several $N^\ast$ resonances 
at low and high $Q^2$ (with $Q^2=-q^2$)~\cite{Aznauryan12,Burkert04,CLAS}. 
This wealth of new experimental data
establishes new challenges for the theoretical models and calculations, 
since,
in the impossibility of solving exact QCD 
in the
momentum transfer regime  of $Q^2=0-10$ GeV$^2$, reliable 
effective and phenomenological approaches are unavoidable.

In this context, we developed a covariant constituent quark model
for the baryons within the spectator framework~\cite{Gross}
for a quark-diquark system~\cite{Nucleon,NDelta,NDeltaD,Omega,ExclusiveR,Nucleon2,NucleonDIS}.   
Because the construction of the electromagnetic current is based
upon the vector meson dominance mechanism, 
it was possible to apply the model also to the lattice QCD regime 
in a domain of unphysical large pion masses~\cite{LatticeD,Omega,Lattice}.  
The model is constrained by the physical data 
for the nucleon and $\gamma^\ast N \to \Delta$ 
data~\cite{Nucleon,NDeltaD,LatticeD}
as well as the $\gamma^\ast N \to \Delta$ lattice QCD data~\cite{LatticeD}. 
The evidence of the predictive power of the model comes from its results 
obtained without further parameter tuning, for the form factors of the
reactions $\gamma^\ast N \to \Delta(1600)$ \cite{Delta1600} 
as well as the reaction $\gamma^\ast N \to N^\ast$ 
where $N^\ast$ can be first radial excitation 
of the nucleon $N^\ast(1440)$, 
and the negative parity partner of the nucleon $N^\ast(1535)$
\cite{Roper,S11,S11scaling}.
Moreover, the extension of the model to the strangeness sector 
was successful in the description 
of the baryon octet~\cite{OctetFF} and 
baryon decuplet form factors~\cite{Omega}. 
All parameters in the model have a straightforward interpretation: 
they give, for instance,
the momentum scales that determine 
the extension of the particle, and the coupling of 
the photon with the constituent quark. 

Importantly, the information extracted from the electromagnetic 
excitation reactions is also relevant for the interpretation 
of production processes induced by strong probes. 
Of particular interest is the study of 
$NN$ collisions in 
elementary nucleon-nucleon reactions 
and in the nuclear medium~\cite{Frohlich10,Zetenyi03,Kaptari06,Kaptari09,Weil12,Agakishiev10,HADES2,Shyam10}.
In this sector, the
HADES experiments of heavy-ion collisions in the 1-2 GeV range  
play a unique 
role in accessing nuclear medium modifications at intermediate 
and high energies~\cite{Weil12,HADES1,Tlusty10}.
Furthermore, in the near future, the FAIR facility 
will expand these experiments further to a higher 
energy regime~\cite{Frohlich10,Kaptari06,Agakishiev10,Shyam10}. 
In both cases, independently of the energy domain under scrutiny, 
the di-lepton channel, the first of which is the low mass di-electron channel, 
is one of the interesting production channels from heavy-ion collisions 
expected to signal in-medium behavior. It is crucial, 
for the interpretation of both present and planned 
di-lepton production data from heavy-ion experiments 
at intermediate energies, to have a reliable baseline made 
of experimental reference from nucleon-nucleon scattering -- 
one of the objectives of the HADES experiments. 

Nonetheless, one also needs  an extension of the knowledge 
gained from the experimental studies 
on elementary electromagnetic transitions, 
to the timelike region ($Q^2=-q^2 < 0$) since
the description of the $NN \to NN e^+ e^-$ reaction 
involves
baryon electromagnetic 
transition form factors in that kinematic regime~\cite{Frohlich10,Kaptari06,Shyam10}. 
Natural requirements of this extension is that it is well constrained, 
appears to be robust when tested by its predictions, 
and allows a direct  physical interpretation of the parameters involved. 

Therefore, in this work we extend our  form factor calculations
for the $\gamma^\ast N \to \Delta$ reaction to the timelike region
and calculate the partial width decays $\Delta \to \gamma N$
and $\Delta \to e^+ e^- N$.
We follow the procedure 
of the standard simulation packages that treat 
the low mass di-electron production data 
as a Dalitz decay following a resonance excitation~\cite{Frohlich10}. 
We note, in particular, that according to 
Ref.~\cite{Frohlich10} the fraction of di-lepton events 
compared to the hadronic channels 
depends significantly on the resonance mass $W$, 
and on the details of the dependence on $q^2$
of the transition form factors, 
two features that call for studies 
as the one we describe here, where such sensitivities are investigated.

We start with the valence quark model presented in Ref.~\cite{LatticeD} 
for the $\gamma^\ast N \to \Delta$ reaction.
That model has two important ingredients:
the contributions from the quark core,
and a contribution of the pion cloud dressing.
In the quark core component,
the  $\Delta$ system has a quark-diquark effective
structure with an S-wave orbital state and 
small D-wave admixtures~\cite{LatticeD,NDelta,NDeltaD}. 
We take here only the dominant S-state 
contribution which is largely responsible for the magnetic dipole 
transition form factor $G_M^\ast$, because 
D-wave states give only small contributions to the 
$\Delta$ wave function ($\le 1\%$) \cite{LatticeD}.
Since the electric and Coulomb quadrupole 
form factors are determined by
the D-state admixture coefficients~\cite{NDeltaD},
in the S-state approximation those two sub-leading form factors
become identically zero. This is a reasonable approximation because they
are  indeed
small when compared with $G_M^\ast$ \cite{NDelta,Pascalutsa07}. 
To the quark core contributions it is necessary to add contributions from the
pion cloud, in order to 
describe the reaction 
in the physical regime for small 
momentum transfer~\cite{NDelta,NDeltaD}.
An important feature of our model is that it describes the $\gamma^\ast N \to \Delta$ 
reaction in the physical and lattice QCD regimes.

The extension of the model to the timelike region presented
here is done directly
by extrapolating the valence quark model~\cite{NDelta},
fixed in the spacelike region, to
the kinematic conditions of the timelike region. 
This means that an arbitrary mass $W$ of the $\Delta$ is taken to
replace its physical mass value.
We also need to generalize the photon-quark current 
to the timelike region, while keeping 
its vector meson dominance parametrization~\cite{Nucleon,NDelta}. 
This is done by 
adding a finite width to the vector meson pole of the current.
As for the pion cloud contributions,
we study two different extensions to the timelike 
region.  Although similar in the spacelike regime~\cite{NDelta,NDeltaD},
they have very different behaviors in the timelike region. 
From the obtained results we conclude that the model 
which includes the $\chi$PT  constraints  is favored.

This work is organized in the following way:
in Sec.~\ref{secFormalism} we introduce 
the formalism that relates the $\Delta$ Dalitz decay 
with the $\gamma^\ast N \to \Delta$ electromagnetic
form factors; in Sec.~\ref{secSQM} the spectator 
quark model is introduced and the 
explicit expressions for the form factors are presented;
in Sec.~\ref{secResults} we show
our results for the decay widths of
$\Delta \to \gamma N$ and $\Delta \to e^+ e^-  N$, and for 
the $\Delta$ mass distribution, as a function 
of $W$; 
finally in Sec.~\ref{secConclusions}
we summarize and draw our conclusions.

\section{Breit-Wigner distribution for the $\Delta$ resonance}
\label{secFormalism}

In the simulations of $NN$ reactions one has to take into 
account the intermediate excitations of the 
nucleon, and the $\Delta$ (spin and isospin 3/2) 
resonance is 
the first relevant one~\cite{Frohlich10,Zetenyi03,Weil12,Kaptari06,Kaptari09,Shyam10}. 
For that purpose we calculated
the contribution of the  $\Delta$(1232) state 
 to the cross section, 
for an arbitrary resonance mass $W$
which can differ 
from the resonance pole (defining the mass $M_\Delta$).
The most usual ansatz is the relativistic 
Breit-Wigner distribution~\cite{Frohlich10,Teis97,Wolf90}, given by
\ba
g_\Delta(W)= A
\frac{W^2 \Gamma_{tot}(W)}{(W^2-M_\Delta^2)^2 + W^2 
\left[ \Gamma_{tot} (W)\right]^2},
\label{eqG}
\ea
where $\Gamma_{tot}$ is the total width,
dependent of $W$,
and $A$ is a normalization factor 
determined by the condition $\int dW g_\Delta(W)=1$.
The total width can be decomposed into the contributions 
from the independent decay channels~\cite{Frohlich10}:
\ba
\Gamma_{tot} (W)= \Gamma_{\pi N}(W) + \Gamma_{\gamma N}(W) +
\Gamma_{e^+ e^- N}(W),
\ea
respectively for the decays 
$\Delta \to \pi N$, $\Delta \to \gamma N$ 
($\gamma$ represents a real photon) and 
$\Delta \to e^+ e^- N$.

The dominant process is the decay $\Delta \to \pi N$,
which can be described by the well known ansatz~\cite{Wolf90,Frohlich10} 
\ba
 \Gamma_{\pi N}(W) = \frac{M_\Delta}{W} 
\left(\frac{q_\pi(W)}{q_\pi(M_\Delta)} \right)^3 
\left(\frac{\nu (W)}{\nu(M_\Delta)} \right)^2 \Gamma_{\pi N}^0,
\label{eqGammaPiN}
\ea
where $q_\pi(W)$ is the pion momentum 
for the decay of a $\Delta$ with mass $W$, 
and  $\Gamma_{\pi N}^0$ is the $\Delta \to \pi N$ 
partial width for the physical $\Delta$ 
[$\Gamma_{\pi N}^0 \equiv \Gamma_{\pi N}(M_\Delta)$].
The function $\nu(W)$ 
is a phenomenological 
function given by 
\ba
\nu(W)=\frac{\beta^2}{\beta^2 + q_\pi^2(W) },
\ea   
where $\beta$ as a cutoff parameter.
Following Refs.~\cite{Frohlich10,Wolf90} we use $\beta=300$ MeV.

As for the components $\Gamma_{\gamma N}(W)$ and 
 $\Gamma_{e^+ e^- N}(W)$, they will be determined by the
$\Delta$ Dalitz decay, as described next.

\subsection{$\Delta$ Dalitz decay}

The $\Delta$ Dalitz decay can be expressed 
in terms of the function $\Gamma_{\gamma^\ast N} (q; W)$,
where  $\gamma^\ast N$ is a short notation for the reaction
$\Delta \to \gamma^\ast N$, and $\gamma^\ast$ 
represents a virtual photon, with squared momentum $q^2 \ge 0$ 
(i.e.~timelike).
The variable $q$ is defined by $q=\sqrt{q^2}$.
The case $q^2=0$ corresponds to the real photon limit.

The    $\Gamma_{\gamma^\ast N} (q; W)$ function 
can be written~\cite{Frohlich10,Krivoruchenko01} as
\ba
\Gamma_{\gamma^\ast N} (q; W)= 
\frac{\alpha}{16} 
\frac{(W+M)^2}{M^2 W^3} 
\sqrt{y_+ y_-} y_-  |G_T(q^2;W)|^2,
\nonumber \\
\label{eqGammaT}
\ea
where $M$ is the nucleon mass, $\alpha \simeq 1/137$ 
the fine-structure constant, and
\ba
y_\pm = (W \pm M)^2- q^2. 
\ea
The function $|G_T(q^2; W)|$ depends on the $\gamma^\ast N \to \Delta$ 
transition form factors: 
$G_M^\ast$ (magnetic dipole), 
$G_E^\ast$ (electric quadrupole) 
and $G_C^\ast$ (Coulomb quadrupole)~\cite{Jones73}, and is given by
\ba
|G_T(q^2; W)|^2 &=&
\left|G_M^\ast(q^2;W)\right|^2 +
3 \left| G_E^\ast(q^2;W) \right|^2 \nonumber \\
& &+
\frac{q^2}{2 W^2}\left| G_C^\ast(q^2;W)\right|^2.
\label{eqGMT}
\ea 
In this equation we note 
that the contribution of each form factor 
will always be real and positive, even if the 
form factors are complex.

Equation (\ref{eqGammaT}) allows the calculation
of any $\Delta \to \gamma^\ast N$ decay,  once a
model for the $\gamma^\ast N \to \Delta$ form factors 
in the timelike region is provided.
Note however that in Eq.~(\ref{eqGMT}) the 
form factors  can be directly measured only for  $W=M_\Delta$.
Consequently, any estimation of the function 
$\Gamma_{\gamma^\ast N} (q; W)$ has to be done 
using models that can be constrained only in the limit $W=M_\Delta$.
The implication is that such models should be largely tested 
for their predictions in other different conditions.
Another detail on Eq.~(\ref{eqGammaT}) 
is that $y_-$ vanishes for $q^2=(W-M)^2$. 
As we discuss later, this point corresponds also 
to the upper limit allowed to $q^2$ for the reaction to occur.

\subsection{Explicit expressions for 
$\Gamma_{\gamma N}(W)$ and $\Gamma_{e^+ e^- N}(W)$}

We present now the expressions for 
$\Gamma_{\gamma N}(W)$ and $\Gamma_{e^+ e^- N}(W)$.
The first function is given by Eq.~(\ref{eqGammaT}) 
for the $q^2=0$ limit~\cite{Frohlich10,Wolf90,Krivoruchenko02}
\ba
\Gamma_{\gamma N}(W)= \Gamma_{\gamma^\ast N} (0; W). 
\label{eqGamma0}
\ea
As for $\Gamma_{e^+ e^- N}(W)$ it will be determined by
integrating 
the function 
\ba
\Gamma_{e^+ e^- N}^\prime(q,W) \equiv 
\frac{d \Gamma_{e^+ e^- N}}{d q} (q;W), 
\label{eqGamma1}
\ea
according with 
\ba
\Gamma_{e^+ e^- N}(W)=
\int_{2m_e}^{W-M} \Gamma_{e^+ e^- N}^\prime(q,W) \, dq
\label{eqInt1}
\ea
Note that the integration holds for the 
interval $4m_e^2 \le q^2 \le (W-M)^2$, where $m_e$ is the 
electron mass.
In this case the lowest squared momentum 
corresponds to $q^2=4 m_e^2$, the minimum 
possible value for a physical  $e^+ e^-$ pair.
The upper limit 
is determined by the maximum value 
for $q^2$ needed for the $\Delta$ with mass $W$ to decay into 
a nucleon (mass $M$) and is discussed in Appendix~\ref{appA}.

The function $\Gamma_{e^+ e^- N}^\prime(q,W)$ 
can be determined~\cite{Frohlich10,Wolf90} by
\ba
\Gamma_{e^+ e^- N}^\prime(q,W)=
 \frac{2 \alpha}{3 \pi q} \Gamma_{\gamma^\ast  N}(q;W).
\label{eqGamma2}
\ea
The function $\Gamma_{e^+ e^- N}^\prime(q,W)$ diverges when $q \to 0$,
due to the presence of $1/q$. 
However, this is not a problem, since in (\ref{eqInt1}) 
the lower limit of the integration
variable $q$ (given by $2m_e$) prevents
the integral from diverging.

To proceed from here, the calculation 
of the partial widths  $\Gamma_{\gamma N}(W)$ and $\Gamma_{e^+ e^- N}(W)$ 
requires a model for the $\gamma^\ast N \to \Delta$ form factors 
in the timelike region, for an arbitrary $\Delta$ mass $W$.
In the past at least three models were proposed 
to this reaction: 
constant form factors~\cite{Frohlich10,Zetenyi03}, 
a two-component quark model 
(model with valence and pion cloud components)~\cite{Frohlich10,Wan05,Iachello04,Bijker04,Wan06}
and 
a vector meson dominance model from Ref.~\cite{Krivoruchenko02}.
In the next section we propose a new model based 
on the spectator formalism.
This model  can be described also as a two-component quark model.
What is specific of our model is that,
in addition to the constraints from the spacelike physical data, 
our model was also constrained by the spacelike 
lattice QCD data \cite{LatticeD}.

\section{Spectator quark model}
\label{secSQM}

We will focus now in the covariant spectator quark model 
for the $\gamma^\ast N \to \Delta$ reaction 
\cite{NDelta,NDeltaD,LatticeD}.
Here, we will describe briefly the  properties
of the model and summarize the important results.
In its simple version, when the nucleon and $\Delta$ 
are both approximated by an S-state configuration 
for the quark-diquark system, the transition 
form factors are restricted to the dominant 
magnetic dipole form factor, and is decomposed~\cite{NDelta} into
\ba
G_M^\ast(Q^2;W)=
G_M^B(Q^2;W) + G_M^\pi(Q^2;W),
\label{eqGMsl}
\ea
where $G_M^B$ is the contribution of the 
quark core and $G_M^\pi$ represents the effect of the pion cloud. 
In the previous equation $W$ 
replaces $M_\Delta$, the physical $\Delta$ mass
used in the previous applications~\cite{NDelta,NDeltaD}.
Because our original model and formulas were developed in the spacelike region,
we maintain here the use of the variable $Q^2$ which stands for $-q^2$. 
Explicitly $G_M^B$ is written as~\cite{NDelta}
\ba
G_M^B(Q^2;W)= \frac{8}{3\sqrt{3}} \frac{M}{M+W} f_v(Q^ 2) {\cal I}(Q^2),
\label{eqGMB}
\ea
where
\ba
{\cal I}(Q^2)= \int_k \psi_\Delta(P_+,k) \psi_N(P_-,k),
\label{eqInt}
\ea
is the overlap integral of the nucleon and $\Delta$ 
radial wave functions which depend on the nucleon ($P_-$), 
the $\Delta$ ($P_+$) and intermediate 
diquark ($k$) momenta. 
The integration sign indicates the covariant 
integration in the diquark momentum $k$: 
$\int_k \equiv \int \frac{d^3 {\bf k}}{(2\pi)^2 2 E_D}$,
where $E_D=\sqrt{m_D^2+ {\bf k}^2}$ is the diquark 
energy ($m_D$ is the diquark mass).
Explicit expressions for the  
nucleon radial wave function $\psi_N$ and 
the $\Delta$ radial wave function
$\psi_\Delta$ will be presented later.
As for the factor $f_v$ it is represented by
\ba
f_v(Q^2)= f_{1-}(Q^2) + \frac{W+M}{2M} f_{2-}(Q^2),
\label{eqFv}
\ea
where $f_{i -}$ ($i=1,2$) are the quark (isovector) form factors, 
that parameterize the electromagnetic 
photon-quark coupling~\cite{Nucleon,NDelta,Omega}.
The $f_{i \pm}$ parameterizations 
will be discussed in more detail in the next subsection.

The pion cloud parametrization $G_M^\pi$ 
was established in the physical regime using the 
factorization~\cite{NDeltaD}
\ba
G_M^\pi(Q^2;W)= 3 \lambda_\pi G_D(Q^2) 
\left( \frac{\Lambda_\pi^2}{\Lambda_\pi^2+ Q^2}\right)^2,
\label{eqGMpi}
\ea
where $G_D=\left(1+ \frac{Q^2}{0.71} \right)^{-2}$,
with $Q^2$ in GeV$^2$, has the usual dipole functional form,
and $\lambda_\pi$ and $\Lambda_\pi$
are parameters that define the strength and the 
falloff of the pion cloud effects. 
In particular we take $\lambda_\pi=0.441$ and $\Lambda_\pi^2=1.53$ GeV$^2$ 
following Refs.~\cite{NDeltaD,LatticeD}.
More details of the model in the physical regime ($W=M_\Delta$) 
can be found in Refs.~\cite{NDelta,NDeltaD}.
Since the pion cloud parameterization 
given by the right-hand-side of Eq.~(\ref{eqGMpi}) has no explicit dependence 
on $M_\Delta$, in its extension to $W \ne M_\Delta$
we consider no explicit  
dependence on $W$ either. 
Then, for  $W \ne M_\Delta$ our choice was to keep
$G_M^\pi$ independent of $W$, that is 
$G_M^\pi(Q^2;W)=G_M^\pi(Q^2;M_\Delta)$, and
the variable $W$ could have been dropped in the
Eq.~(\ref{eqGMpi}).

A general comment about the decomposition (\ref{eqGMsl}) is in order.
In the spectator framework 
the component $G_M^B$, given by Eq.~(\ref{eqGMB}), 
is limited by the condition ${\cal I}(0) \le 1$, which follows from
the normalization of the nucleon and 
$\Delta$ radial wave functions and the 
Cauchy-Schwartz-H\"{o}lder inequality. 
This implies that $G_M^B(0,M_\Delta) \le 2.07$ \cite{NDelta}.
Since the experimental value is  
$G_M^\ast(0,M_\Delta) \simeq 3$ it follows that 
the description of the reaction near $Q^2=0$ 
is not possible, unless the 
contribution of the pion cloud is significant:
more than $30\%$ of the total result.
The underestimation of $G_M^\ast(0,M_\Delta)$ 
is a  result common to several
models based on constituent quark 
degrees of freedom alone~\cite{NDelta}.

\subsection{Quark current}

In the spectator quark model the electromagnetic interaction 
with the quarks is represented in terms 
of Dirac and Pauli electromagnetic form factors,  $f_{1\pm}$ (Dirac)
and $f_{2\pm}$ respectively,  for the quarks~\cite{Nucleon,NDelta,Omega}. 
Using the vector meson dominance (VMD) mechanism,
those form factors are 
parametrized  as
\ba
& &
f_{1\pm}(q^2)=\lambda_q
+ (1-\lambda_q) \frac{m_v^2}{m_v^2-q^2} -
c_\pm \frac{M_h^2 q^2}{(M_h^2-q^2)^2} \nonumber \\
& &
f_{2\pm}(q^2)=
\kappa_\pm \left\{ d_\pm
\frac{m_v^2}{m_v^2-q^2} 
+(1-d_\pm) \frac{M_h^2}{M_h^2-q^2}
\right\}, \nonumber \\
& &
\label{eqQcurrent}
\ea
where $m_v$ is a light vector meson mass, $M_h$ 
is a mass of an effective heavy vector meson,
$\kappa_\pm$ are quark anomalous magnetic moments,
$c_\pm,d_\pm$ are mixture coefficients and  $\lambda_q$ 
a parameter related with the quark density number in 
deep inelastic scattering~\cite{Nucleon}.
In the applications we take $m_v= m_\rho$ ($\simeq m_\omega$), to
include the physics associated with the $\rho$-meson,
and $M_h= 2 M$ (twice the nucleon mass) for
effects of meson resonances with a larger mass than the $\rho$.
Note that both functions $f_{1-}$ and $f_{2-}$ have a pole at $q^2=m_\rho$ and at $q^2= M_h^2$.
Hereafter we will refer to these poles as
$\rho$-poles and $M_h$ poles, respectively.

The parametrization (\ref{eqQcurrent}) is particularly 
useful for applications of the model to the 
lattice QCD spacelike regime.
In fact, the decomposition of the current into contributions from the
vector meson poles ($m_\rho$ and $M_h=2M$) is very convenient 
for a extension of the model to a regime 
where those poles can be replaced by the $m_\rho$ and $M$ values
given by the lattice calculations, without 
introducing any additional parameters.
Examples of successful applications to the lattice regime can be found 
in Refs.~\cite{Lattice,LatticeD,Omega,OctetFF}.
In Refs.~\cite{Lattice,LatticeD}, in particular, one can see how
well the model describes the lattice data from 
Ref.~\cite{Alexandrou08} for the $\gamma^\ast N \to \Delta$ 
reaction, particularly for pion masses $m_\pi > 400$ MeV
where the pion cloud effects $G_M^\pi$ are suppressed.
The valence quark contribution~\cite{Lattice,NDeltaD} is 
also compatible with the estimation 
of the bare contribution from the EBAC model~\cite{Diaz07}.
The successful description of the $G_M^\ast$ lattice data 
shows that the valence quark calibration of our model 
is under control.

To stress  the first problem of the extension 
of (\ref{eqQcurrent}) to the case $q^2>0$, 
in Eq.~(\ref{eqQcurrent}), we used explicitly 
the variable $q^2$ instead of the variable $Q^2$ 
employed in Refs.~\cite{Nucleon,NDelta,NDeltaD}:
singularities appear at $q^2=m_\rho^2$ and $q^2= M_h^2$.
The larger poles are not problematic
for moderated $W$,
since as shown in Appendix~\ref{appA}, $q^2 \le (W-M)^2$.  But 
the case  $q^2=m_\rho^2$ has to be taken with care.
Such pole is a consequence  
of having the  $\rho$ meson as a stable particle, with a zero mass width.
One can overcome this  limitation by introducing 
a finite width $\Gamma_\rho$ in the $\rho$-propagator 
$m_\rho^2/(m_\rho^2-q^2)$, with the replacement 
$m_\rho \to m_\rho -\sfrac{i}{2}\Gamma_\rho$.
A non-zero width $\Gamma_\rho$ leads then 
to the substitution
\ba
\frac{m_v^2}{m_v^2 -q^2} 
& \to & 
\frac{m_\rho^2}{m_\rho^2 -q^2 -i m_\rho  \Gamma_\rho}  \nonumber \\
& \to & 
\frac{m_\rho^2\left[ (m_\rho^2- q^2) + i m_\rho \Gamma_\rho\right]}{
(m_\rho^2- q^2)^2 + m_\rho^2 \Gamma_\rho^2}. 
\label{eqVMD}
\ea
Note that this procedure 
induces an imaginary part in the bare quark contributions 
for the form factors.
The $\rho$-width $\Gamma_\rho$ is in fact 
a real function of $q^2$ defined only 
for $q^2 > 0$, as we discuss next, and
therefore the results in the spacelike 
regime are unaffected by the redefinition (\ref{eqVMD}).

The $\rho$ width can be measured only for the physical 
decay of the $\rho$, when $q^2=m_\rho^2$.
The experimental value is $\Gamma_\rho^0=\Gamma_\rho (m_\rho^2)=0.149$ GeV (PDG) \cite{PDG}.
For $q^2\ge 0$
one has to consider some parametrization 
for $\Gamma_\rho (q^2)$.
An usual parametrization is~\cite{Connell95,Connell97,Gounaris68}: 
\ba
\Gamma_\rho(q^2)= \Gamma_\rho^0 \left( 
\frac{q^2-4 m_\pi^2}{m_\rho^2-4 m_\pi^2}\right)^{3/2}
\frac{m_\rho}{q} \theta(q^2- 4m_\pi^2),
\label{eqGammaR}
\ea
where $m_\pi$ is the pion mass 
and $\theta(x)$ the Heaviside step function 
that cuts the contributions for $q^2 \le 4 m_\pi^2$,
below the $2 \pi$ creation threshold (decay $\rho \to 2 \pi$).
The previous formula includes
then the creation of $\pi \pi$ states 
from an off-mass-shell $\rho$.
Equation~(\ref{eqGammaR})
assures that there is no width near $q^2=0$. 
Therefore the imaginary contribution 
appears only for $q^2 > 4 m_\pi^2 \simeq 0.076$ GeV$^2$.

\subsection{Scalar wave functions}

The radial (or scalar) wave functions 
taken in this work, respectively 
for the nucleon and $\Delta$, are
\ba
& &
\psi_N(P,k)= \frac{N_N}{
m_D (\beta_1 + \chi_N) (\beta_2 + \chi_N)} \\
& &
\psi_\Delta(P,k)= \frac{N_\Delta}{
m_D (\alpha + \chi_\Delta)^3},
\ea
where $m_D$ is the diquark mass, $\beta_1,\beta_2$ and $\alpha$ 
are momentum range parameters (in units $m_D$) and
\ba
\chi_B= \frac{(M_B-m_D)^2-(P-k)^2}{M_B m_D},
\label{eqChi}
\ea
for $B=N$ ($M_B=M$) and $B=\Delta$ ($M_B=M_\Delta$),
is a variable without dimensions that includes the 
dependence in the quark momentum $(P-k)^2$.
As for $N_B$, ($B=N,\Delta$) they are positive normalization constants.
See Refs.~\cite{Nucleon,NDelta} for details.
The representation of the wave function in
terms of $\chi_B$ given by Eq.~(\ref{eqChi}) has 
advantages in the applications to the lattice regime~\cite{Lattice,LatticeD,Omega}.

The scalar wave functions are important for the present calculations
because they are part of the overlap integral
defined by Eq.~(\ref{eqInt}).
To apply the expressions to the timelike 
region one has to choose a configuration with $Q^2<0$ ($q^2 >0$). 
That can be achieved by considering the reaction $\gamma^\ast N \to \Delta$ 
in the $\Delta$ rest frame, with the following 
configuration: $P_+=(W,{\bf 0})$ as the $\Delta$ momentum 
and $P_-=(E_N,-{\bf q})$, with $E_N=\sqrt{M^2+{\bf q}^2}$, 
as the nucleon momentum.
In those conditions the photon momentum is represented by
$q=P_+-P_-$, as $q=(\omega, {\bf q})$, 
where
\ba
& &
\omega =\frac{W^2- M^2 + q^2}{2W} \nonumber \\
& &
{\bf q}^2= \frac{(W^2+ M^2-q^2)^2}{4 W^2} - M^2. 
\label{eqPTL}
\ea
Those variables correspond to the timelike region 
when $0 \le q^2 \le (W-M)^2$.
See details in Appendix~\ref{appA}.

Using  Eq.~(\ref{eqGMB}) and the integral (\ref{eqInt}) 
for the kinematics (\ref{eqPTL}),
together with the extension of the current given by (\ref{eqVMD}), 
one can calculate the  contribution for $G_M^B$.
Note that as the function (\ref{eqVMD})
has an imaginary component, 
$G_M^B$ is now complex.

\subsection{Pion cloud contribution}

The most phenomenological part 
of the model presented here is the parametrization 
of the pion cloud contribution through Eq.~(\ref{eqGMpi}).
Although the valence quark parametrization 
has been validated by lattice QCD simulations 
and the EBAC estimations of the quark core 
contributions~\cite{Alexandrou08,Diaz07},
the contributions from the pion cloud 
were estimated only phenomenologically.
In fact they were extracted directly from the physical data, 
after the calibration of the valence quark effects \cite{NDelta,NDeltaD}.

For the pion cloud component of the form factor
we will compare two different generalizations 
of  Eq.~(\ref{eqGMpi}) for the timelike region.
We start with a simple model,  
a naive generalization of the model 
from Refs.~\cite{NDelta,NDeltaD,LatticeD}
to the timelike region.
Next we discuss the possible limitations 
of that approach and introduce a different parametrization 
motivated by the expressions for the 
pion cloud derived from $\chi$PT.

\subsubsection{Naive model (model 1)}

In a first approach we took 
the pion cloud contributions for the 
$G_M^\ast$ form factor by Eq.~(\ref{eqGMpi}),
as in the spacelike regime,
but now evaluated in the timelike kinematic region.
We have to take into consideration now 
the poles for $q^2 > 0$ ($Q^2< 0$).
We re-write $G_D$ as
\ba
G_D(q^2)= \left( \frac{\Lambda_D^2}{\Lambda_D^2-q^2} \right)^2,
\ea
where $\Lambda_D^2=0.71$ GeV$^2$ is the cutoff 
of the dipole form factor.
As it happens to the $\rho-$term in the quark current, also 
this factor has a pole at $q^2=0.71$ GeV$^2$,
but in this case it is a double pole. 
We apply the procedure used before 
to the $\rho$ propagator, i.e.~definiting  
a width to the function $G_D$, by making
\ba
G_D (q^2)&\to& 
\left[ 
\frac{\Lambda_D^2}{(\Lambda_D^2 -q^2)^2+ \Lambda_D^2 \Gamma^2_D}
\right]^2 \times  \label{eqGD2}\\
& &
\left[ 
(\Lambda_D^2-q^2)^2 -\Lambda_D^2 \Gamma^2_D 
+ i 2(\Lambda_D^2-q^2) \Lambda_D \Gamma_D
\right], \nonumber
\ea
where $\Gamma_D$ is the width associated with its pole.
As the poles $m_\rho^2$ and $\Lambda_D^2$ are close 
($q^2 \simeq 0.6$ GeV$^2$ versus $q^2 \simeq 0.7$ GeV$^2$), 
we will use $\Gamma_D(q^2) = \Gamma_\rho(q^2)$.
With $G_D$ defined  as above,
also $G_M^\pi$ is a complex function 
in the timelike regime.

As for the extra dipole factor in Eq.~(\ref{eqGMpi}):
$\left( \frac{\Lambda_\pi^2}{\Lambda_\pi^2-q^2} \right)^2$,
where in the applications $\Lambda_\pi^2 \simeq 1.5 $ GeV$^2$,
it is far way from the $\rho-$poles region.
For $\Delta$ masses not very large 
compared with $M_\Delta$ the possible effect of the finite width 
is less significant since $q^2 < (W-M)^2 < \Lambda_\pi^2$.

\begin{figure}[t]
\vspace{.3cm}
\centerline{
\mbox{
\includegraphics[width=3.0in]{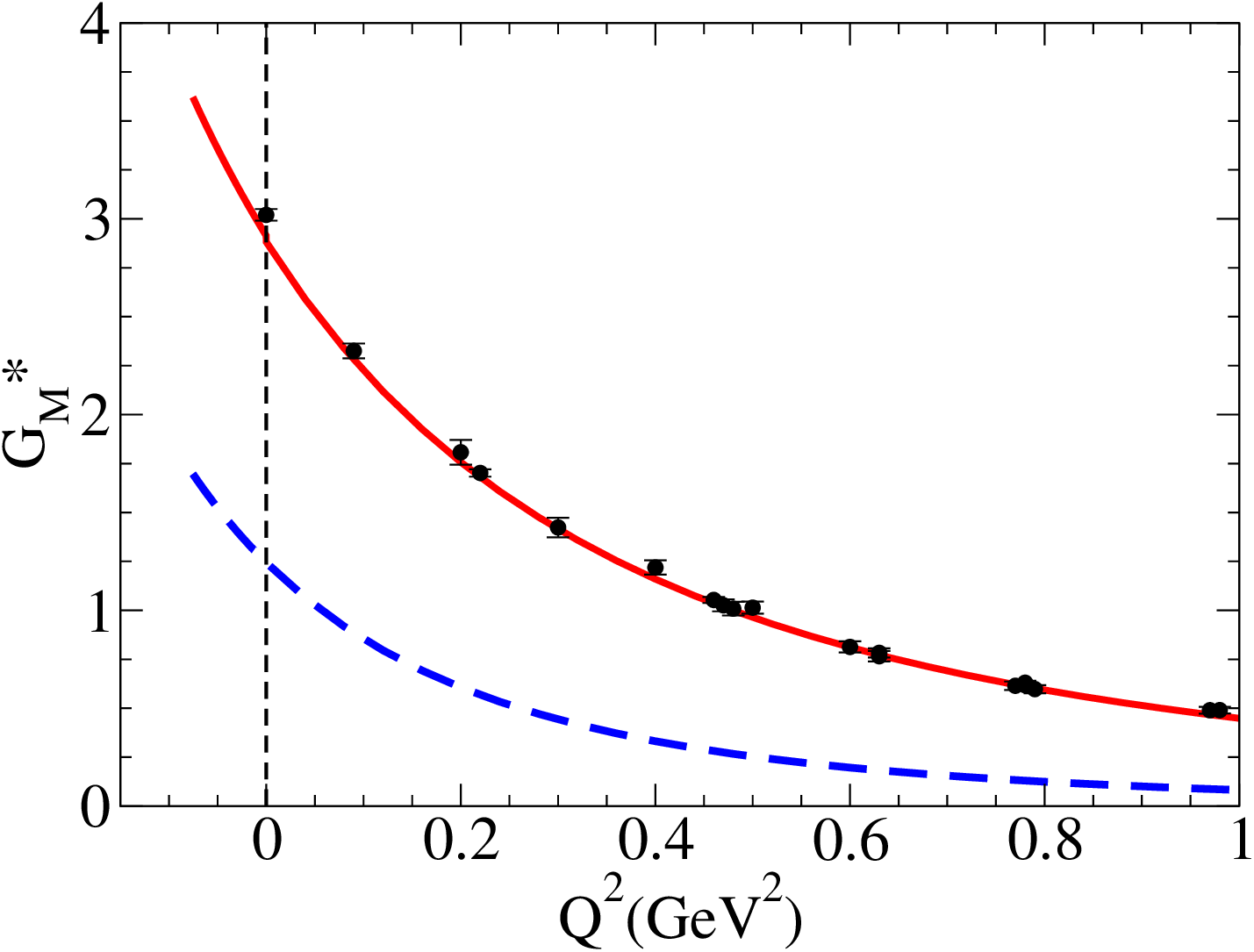}
}}
\caption{\footnotesize{
$\gamma^\ast N \to \Delta(1232)$ magnetic form factor 
in the timelike and spacelike region given by 
the model of Ref.~\cite{LatticeD}.
The data for $Q^2>0$ is the same as the one 
presented in Ref.~\cite{LatticeD}.
The data for $Q^2=0$ is from Ref.~\cite{Tiator01}.
The dashed line represents the contributions 
of the valence quark core (bare contribution).
The solid line is the result of the sum 
of valence quarks and pion cloud contributions.
}}
\label{figGM}
\end{figure}

We call the model defined by Eqs.~(\ref{eqGMpi}) and (\ref{eqGD2})
model 1.
The result of the extension of the model to the 
timelike region for the case $W=M_\Delta$
and $\Gamma_\rho \equiv 0$ is presented in Fig.~\ref{figGM}.
We used the parametrization of Ref.~\cite{LatticeD}
for the valence quark contributions, 
but neglected the D-state contributions ($\le 1\%$).
As in this case $Q^2 \ge -(M_\Delta-M)^2 \simeq -0.086$ GeV$^2$, 
and therefore the allowed values for $Q^2$ 
are far way from the $Q^2<0$ poles,  
the corrections due to the imaginary components are small.
In the figure we show also physical data for 
$Q^2 > 0$ and the result for $Q^2=0$ from Ref.~\cite{Tiator01}.


\subsubsection{$\chi$PT motivated model (model 2)}

Instead of Eq.~(\ref{eqGMpi}) for the pion cloud effect we can use
a different parametrization, based 
in a different combination of multipole functions.
For instance, in the two-component model
from Refs.~\cite{Frohlich10,Wan06} 
the contribution from the pion cloud 
is proportional to the function $F_\rho$, 
interpreted as the $\rho-$propagator, derived from $\chi$PT.
This function $F_\rho$ was 
presented in Refs.~\cite{Iachello73,Wan06,Iachello04} 
taking into account 
the pion loop contributions to the
$\rho-$propagator.
Here we simplified the exact expression in those 
references by assuming its limit when
$q^2 \gg 4 m_\pi^2$, and using the normalization $F_\rho(0)=1$. 
For $Q^2=-q^2 > 0$ we obtained then
\ba
F_\rho (q^2) \simeq
 \frac{m_\rho^2}{m_\rho^2 + Q^2 +
\sfrac{1}{\pi} \frac{\Gamma_\rho^0}{m_\pi} Q^2 
\log \frac{Q^2}{m_\pi^2} }.
\label{eqVMD2}
\ea
In the previous equation,  the physical  
$\rho$-width $\Gamma_\rho^0$ was taken to be $\Gamma_\rho^0=0.149$ GeV.
Reference~\cite{Frohlich10}, uses instead $\Gamma_\rho^0=0.112$ GeV.


\begin{figure}[t]
\vspace{.45cm}
\centerline{
\mbox{
\includegraphics[width=3.0in]{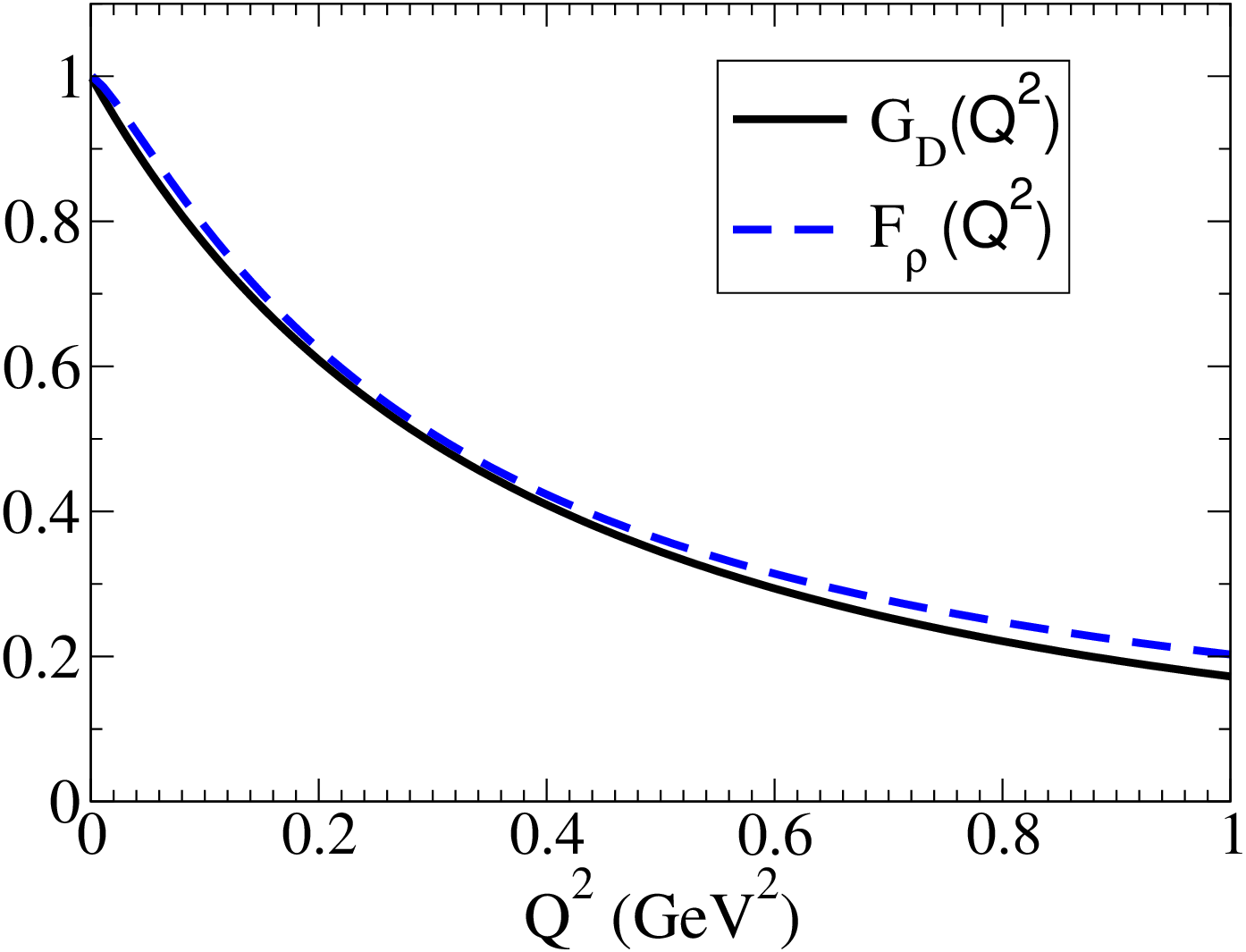}
}}
\caption{\footnotesize{
Comparing the dipole form factor $G_D$
with the function $F_\rho(Q^2)$ given by Eq.~(\ref{eqVMD2}). }}
\label{figFrho}
\end{figure}

\begin{figure*}[t]
\vspace{.3cm}
\centerline{
\mbox{
\includegraphics[width=3.0in]{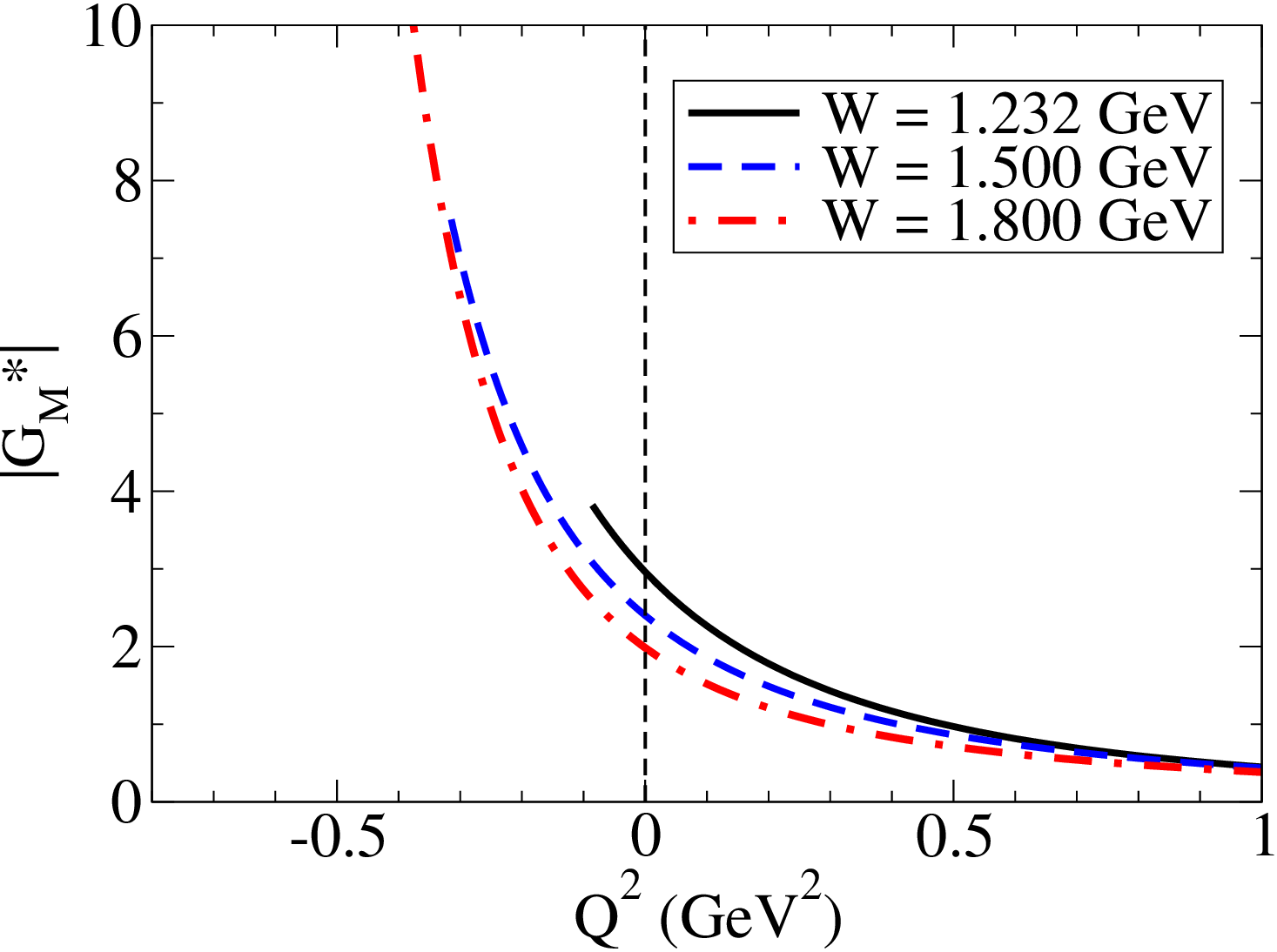} \hspace{1.cm}
\includegraphics[width=3.0in]{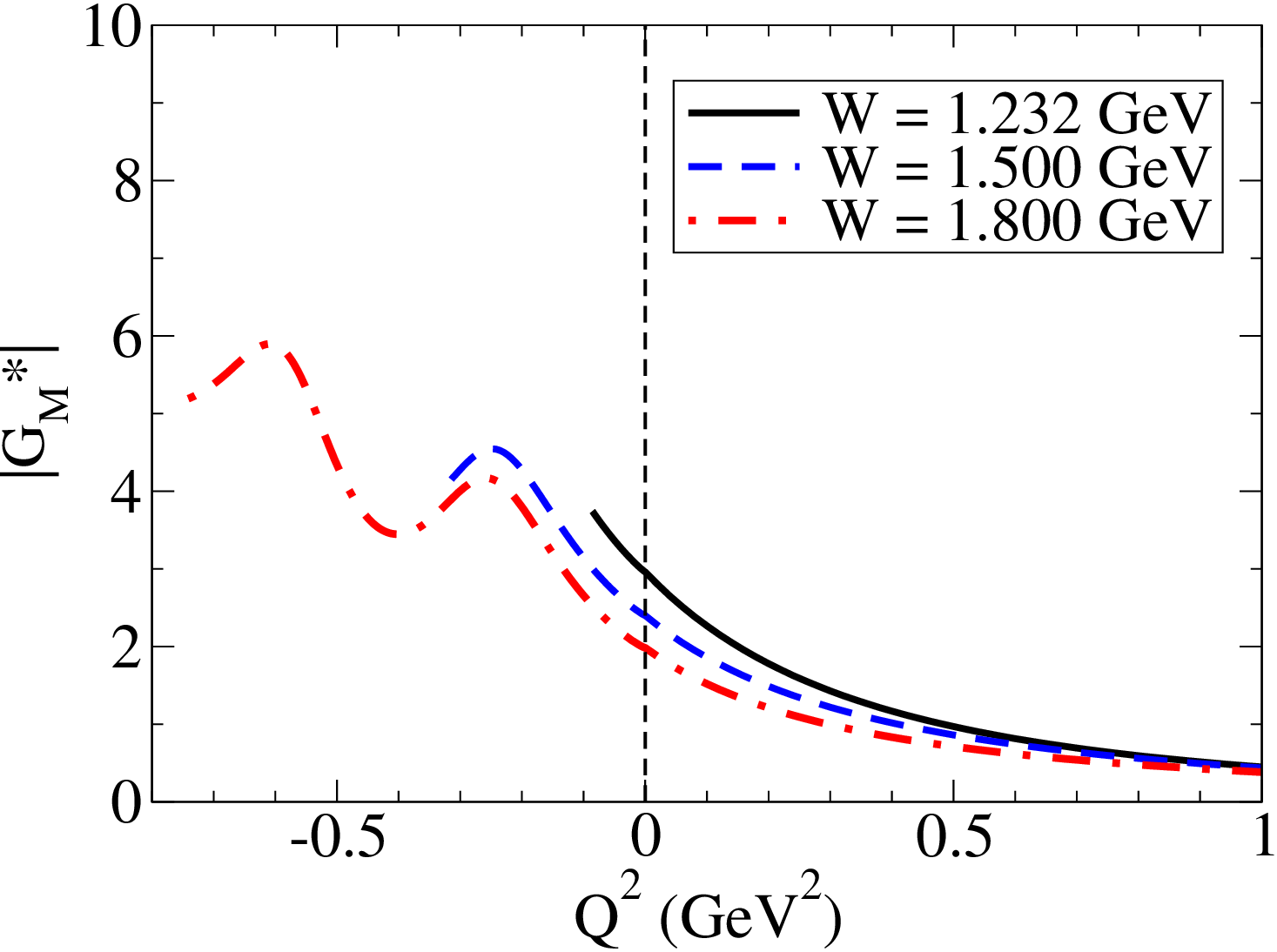}
}}
\caption{\footnotesize{
Results for the modulus of the 
$\gamma N \to \Delta$ magnetic form factor 
for different $\Delta$ masses ($W$).
At left: model 1; at right: model 2.
In both cases the valence quark model 
is from Ref.~\cite{LatticeD}.}}
\label{figGM2}
\end{figure*}

Equation (\ref{eqVMD2}), derived in 
the low $q^2$ chiral perturbation regime, has a faster falloff 
for the $\rho$-propagator [with  $1/(Q^2 \log \frac{Q^2}{m_\pi^2})$] 
than model 1 [with $1/Q^2$] for large $Q^2$. We used it here to explore 
alternative parametrizations to model 1 for the pion cloud contributions.
The parametrization for the pion cloud 
contribution (\ref{eqGMpi}) 
is proportional to the
dipole factor 
$\left( \frac{\Lambda_\pi^2}{\Lambda_\pi^2+ Q^2}\right)^2$,
where $\Lambda_\pi$ is a large cutoff, and
also to $G_D$, the dipole form factor.  
Although the dipole factor depending on $\Lambda_\pi$
was chosen phenomenologically 
and determined by a fit to the data,
one has no reason a priori to 
use the particular form of $G_D$
to parameterize an extra falloff\footnote{ 
The function $G_D$ 
 provides a good 
approximation for the behavior of the nucleon 
electromagnetic form factor at low $Q^2$.} of $G_M^\pi$. 
The inclusion of $G_D$
was motivated by the traditional 
convention of dividing the form factor $G_M^\ast$ 
by $3G_D$ when showing results.
With the parametrization (\ref{eqGMpi})
one has an asymptotic dependence of $G_M^\pi \propto 1/Q^8$.

\begin{figure}[t]
\vspace{.3cm}
\centerline{
\mbox{
\includegraphics[width=2.8in]{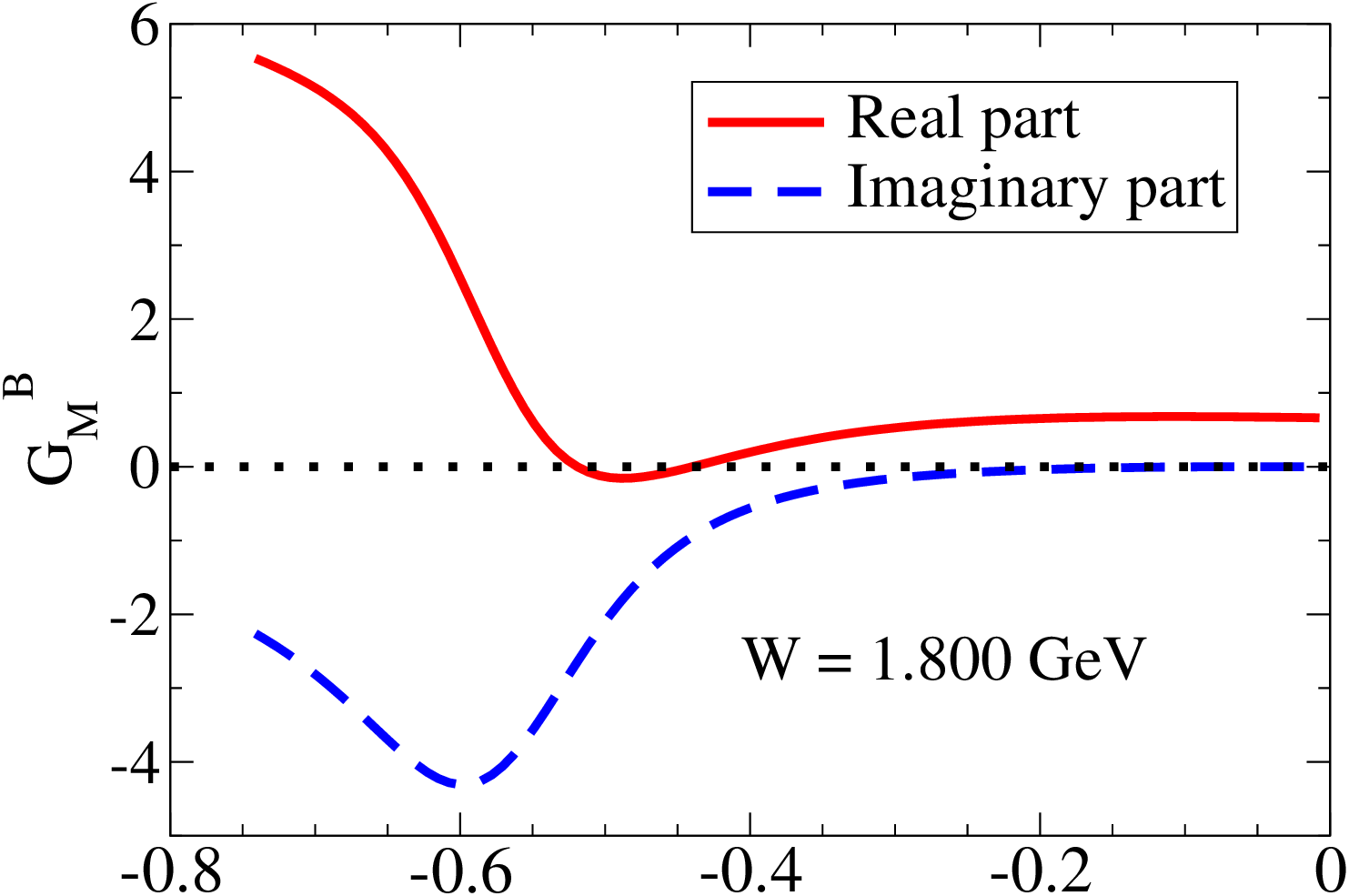}
}}
\centerline{
\vspace{.4cm} }
\centerline{
\mbox{
\includegraphics[width=2.8in]{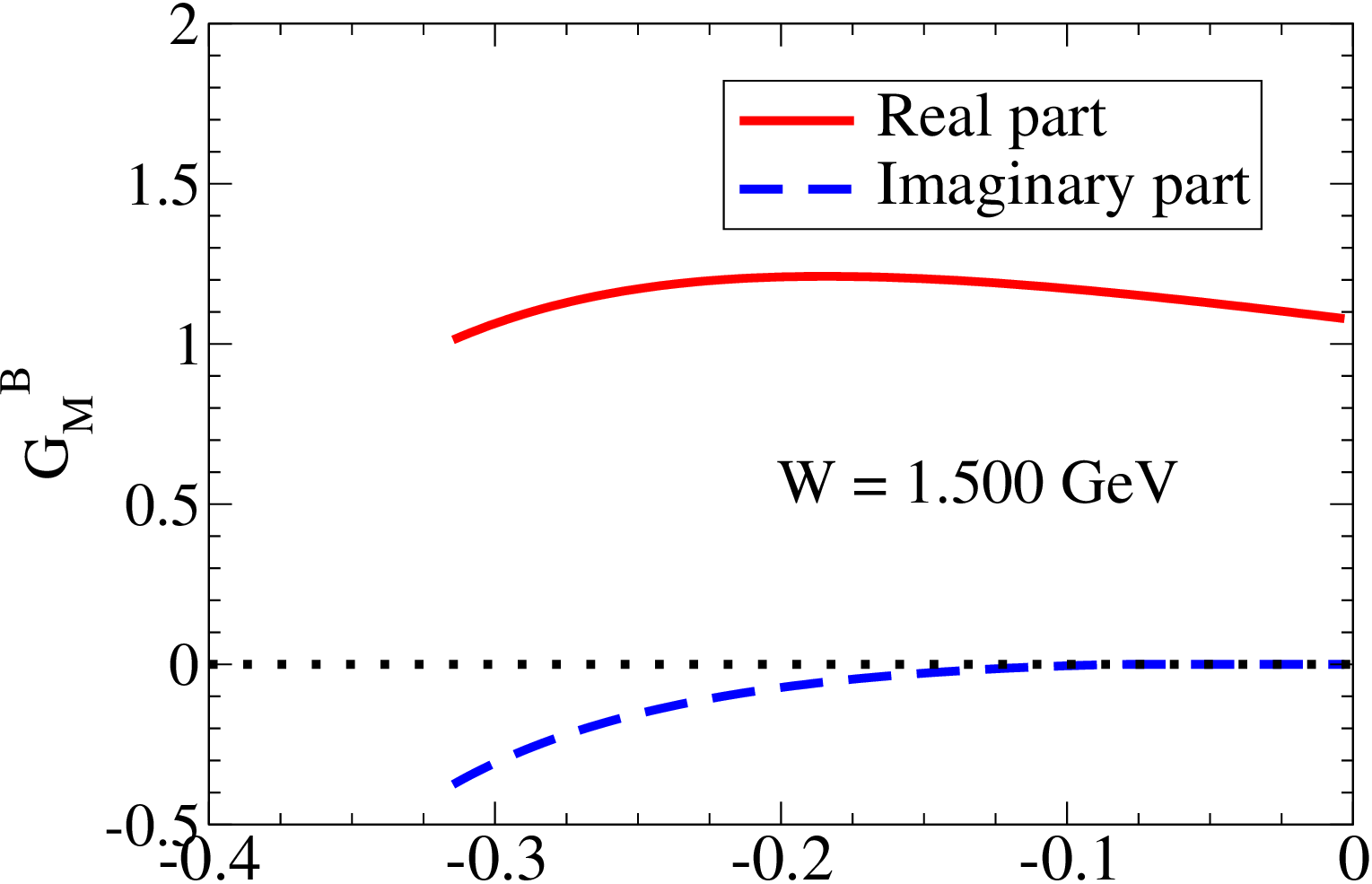}
}}
\centerline{
\vspace{.4cm} }
\centerline{
\mbox{
\includegraphics[width=2.8in]{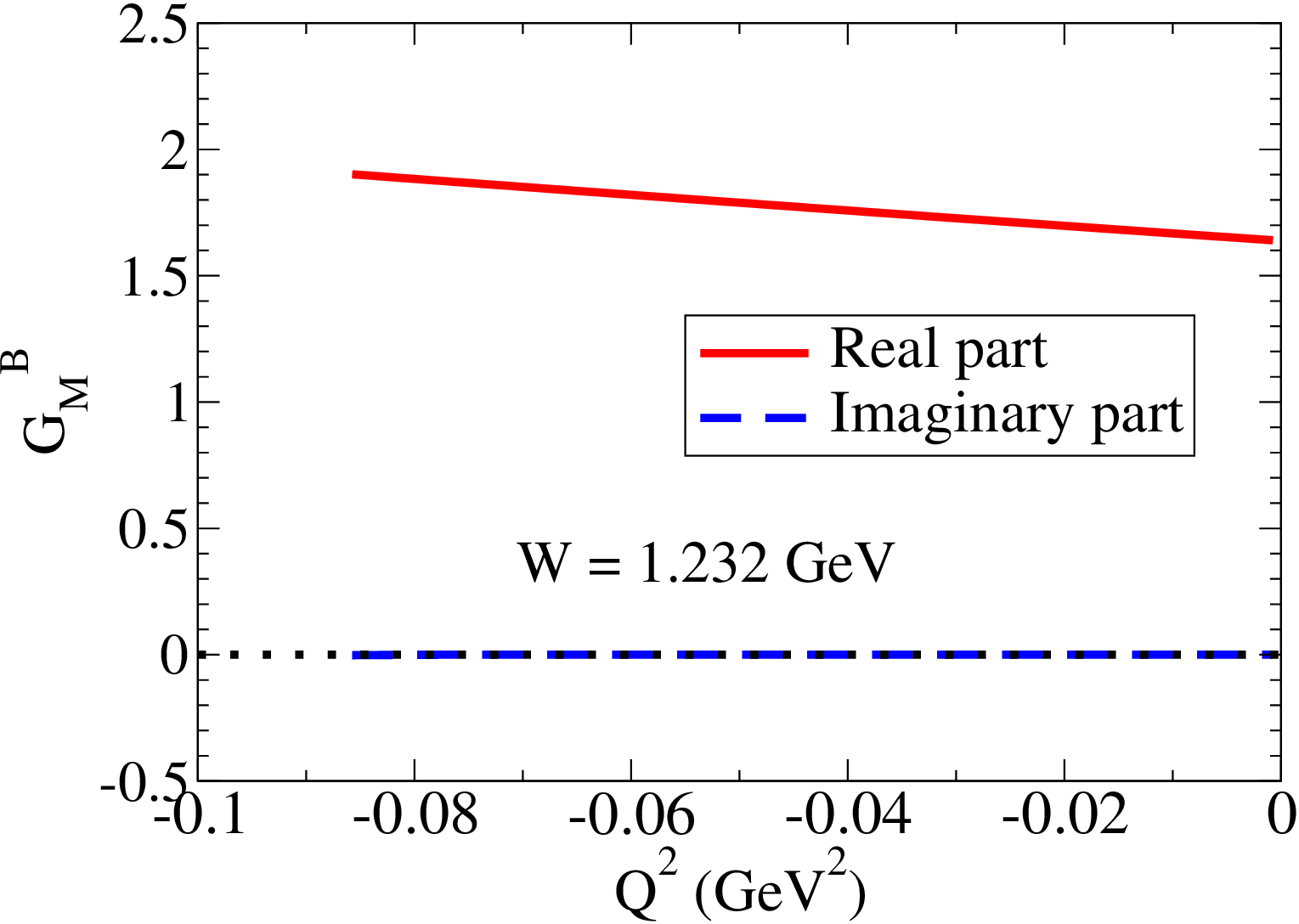}
}}
\caption{\footnotesize{
Valence contribution for 
$\gamma N \to \Delta(1232)$ magnetic form factor 
for $W=1.232, 1.500$ and 1.800 GeV.}}
\label{figGMB}
\end{figure}

We note that in the spacelike region where the pion cloud effects 
are more important, $0< Q^2 < 1$ GeV$^2$,
the functions $G_D$ and $F_\rho$ give very similar results, as seen in
Fig.~\ref{figFrho}.
This suggests that one can also use 
\ba
G_M^\pi (Q^2)= 3\lambda_\pi F_\rho (q^2) 
\left( \frac{\Lambda_\pi^2}{\Lambda_\pi^2 - q^2}\right)^2,
\label{eqGMpi2}
\ea 
with
\ba
F_\rho (q^2) =
 \frac{m_\rho^2}{m_\rho^2 -q^2 -
\sfrac{1}{\pi} \frac{\Gamma_\rho^0}{m_\pi} q^2 
\log \frac{q^2}{m_\pi^2} 
+ i \frac{\Gamma_\rho^0}{m_\pi} q^2} ,
\label{eqVMD2P}
\ea
to extend Eq.~(\ref{eqVMD2})  
to the timelike kinematics.\footnote{
In the transformation from Eq.~(\ref{eqVMD2}) 
to Eq.~(\ref{eqVMD2P}) 
there is an ambiguity from the 
factor 
\ba
\log(-1)= i \pi +  i(2\pi) t,
\nonumber
\ea
where $t$ is an integer.
In this case the  ambiguity is 
fixed by the sign of the imaginary part 
from Ref.~\cite{Frohlich10}.}  
The imaginary part in Eq.~(\ref{eqVMD2P})
is a consequence of  two pion production 
(or transition $\rho \to 2\pi$) which is possible 
in the timelike region when $q^2 \ge 4 m_\pi^2$.

We will call the model defined by Eq.~(\ref{eqGMpi2}) model 2.
One implication of the new form for the pion cloud 
contributions is a falloff  as $1/(q^6 \log q^2)$,
slower than for model 1 
\mbox{[$1/q^8$ falloff].}
[Note that $G_D \propto 1/q^4$ and $F_\rho \propto 1/(q^2 \log q^2)$]

Another important feature of the function (\ref{eqGMpi2})
is that its imaginary part
does not peak for $q^2 = m_\rho^2 \simeq 0.6$ GeV$^2$ 
but for  $q^2 \simeq 0.3$ GeV$^2$, because of the 
logarithm corrections. That effect changes the $Q^2$ dependence 
of the pion cloud contributions to $G_M^\ast$, relatively to model 1.

A note about $F_\rho(q^2)$ given by Eq.~(\ref{eqVMD2P}):
it was derived from the exact result in 
Refs.~\cite{Frohlich10,Iachello73,Iachello04}
in the limit $q^2 \gg 4 m_\pi^2 \simeq 0.08$ GeV$^2$, as explained before.
However, we have checked that our simplified formula, 
although approximate, does not deviate too much from the exact one, 
even when that limit does not hold.
Therefore our formula gives a good qualitative description 
of the chiral behavior in the whole domain $q^2>0$.
We note only that the
results from Eq.~(\ref{eqVMD2P}) 
and the results in Refs.~\cite{Frohlich10,Iachello73,Iachello04} differ
in a slight deviation of the location of the peak 
of the imaginary part of $F_\rho$.
Using Eq.~(\ref{eqVMD2P}) the peak is at $q^2\simeq 0.3$ GeV$^2$, 
while in Ref.~\cite{Frohlich10,Iachello73,Iachello04} 
the peak is at $q^2\simeq 0.4$ GeV$^2$.

Finally, as for the quark current (\ref{eqQcurrent}) 
in the bare quark contributions, we will not replace our parametrization 
of the $\rho$-propagator (\ref{eqVMD}) 
by (\ref{eqVMD2}) since they differ substantially 
and our parametrization  was already 
calibrated by the physical data \cite{Nucleon,NDelta,NDeltaD} 
and lattice data \cite{Lattice,LatticeD}
for the nucleon and $\Delta(1232)$ systems,
in the spacelike region.
A different parametrization for timelike and spacelike regions 
would be inconsistent. 

\section{Results}
\label{secResults}

We will divide the presentation of our results into two parts. 
In the first part we show the results 
of the magnetic form factor $G_M^\ast$, calculated with the 
two models described in the previous section.
In the second part we show the results for
the width functions $\Gamma_{\gamma N}(W)$ and
$\Gamma_{e^+ e^- N}(W)$, and also for the 
$\Delta$ mass distribution function $g_{\Delta}(W)$. 

\subsection{Form factors in the timelike region}

Contrarily to what happens in the spacelike domain,
in the timelike region the form factor $G_M^\ast$ 
has a non-zero imaginary part.
Because of Eq.~(\ref{eqGMT})
we are interested in the absolute 
value of the form factor, 
$|G_M^\ast|$, which enters into $|G_T(q^2; W)|^2$.
Although the form factors are defined for any value of $W \ge M$, 
we show here results for selected  
values of $W$ only. 
We recall
that the range of $Q^2$ for the function
$G_M^\ast(Q^2;W)$  depends on $W$,  as established by the condition 
\mbox{$Q^2 \ge -(W-M)^2$.}

The results for 
$|G_M^\ast|$ at energies 
$W=1.232, 1.500$ and 1.800 GeV, for both models,
are presented in Fig.~\ref{figGM2}.
One notes that the value 
of $|G_M^\ast|$ near $Q^2=0$ decreases with $W$.
We will see that this is a consequence of the 
valence quark contribution given by Eq.~(\ref{eqGMB}). The same effect
was observed in lattice QCD simulations where large pion masses induce
large nucleon and $\Delta$ masses \cite{LatticeD,Alexandrou08}.
In the figure it is also clear
that the two models differ substantially in the
$Q^2$ dependence of $|G_M^\ast|$.
For model 1, $|G_M^\ast|$ increases 
as $Q^2$ decreases, for all values of $W$,
and this behavior is enhanced as $W$ becomes larger. 
A peak (not shown in the graph because it is too large) 
is present, near $q^2=\Lambda_D^2 \simeq 0.71$ GeV$^2$ when $W=1.800$ GeV.
A first conclusion is therefore that model 1 
generates very strong, and probably 
unphysical contributions to $|G_M^\ast|$
in the timelike region. 
Model 2, in contrast, 
gives moderated contributions 
only (larger than in the spacelike region but with the 
same magnitude) and is therefore a much more reasonable model.
To better compare the two models we have to analyze the real 
and imaginary parts of  $G_M^\ast$ separately.

Since the valence quark contributions are common to both models,
we start by looking to the bare term $G_M^B$. 
The results are presented in Fig.~\ref{figGMB}.
For $Q^2 \approx 0$ and values of $W$ not too large when
compared with $W=1.232$ GeV, 
the real part  
dominates, as expected from 
the results for the physical case 
($W=M_\Delta$). 
For larger values of $W$ and low $Q^2$ the real part dominates 
increasingly less.
As for the imaginary part,  we recall that is zero 
down to \mbox{$Q^2 \simeq -0.08$ GeV$^2$} (because $\Gamma_\rho=0$).
But as $Q^2$ decreases, for larger $W$ values, we can observe
the effect 
of the $\rho$-mass poles emerging
at $Q^2= -m_\rho^2 \simeq - 0.6$ GeV$^2$,
and strengthening the imaginary parts of $G_M^B$.
For $W=1.800$ GeV the real part
also increases in the region $Q^2 <  - m_\rho^2$
due  to the impact of the $\rho$-mass poles.
In this case, however, the other terms from 
the VMD parametrization (\ref{eqQcurrent}), 
the constant term and the $M_h$-mass poles, 
are relevant as well.
All these contributions to $G_M^B$ 
are balanced, and therefore reduced, in the final result,
by the pion contributions  $G_M^\pi$ to be discussed next.

\begin{figure*}[t]
\vspace{.3cm}
\centerline{
\mbox{
\includegraphics[width=2.8in]{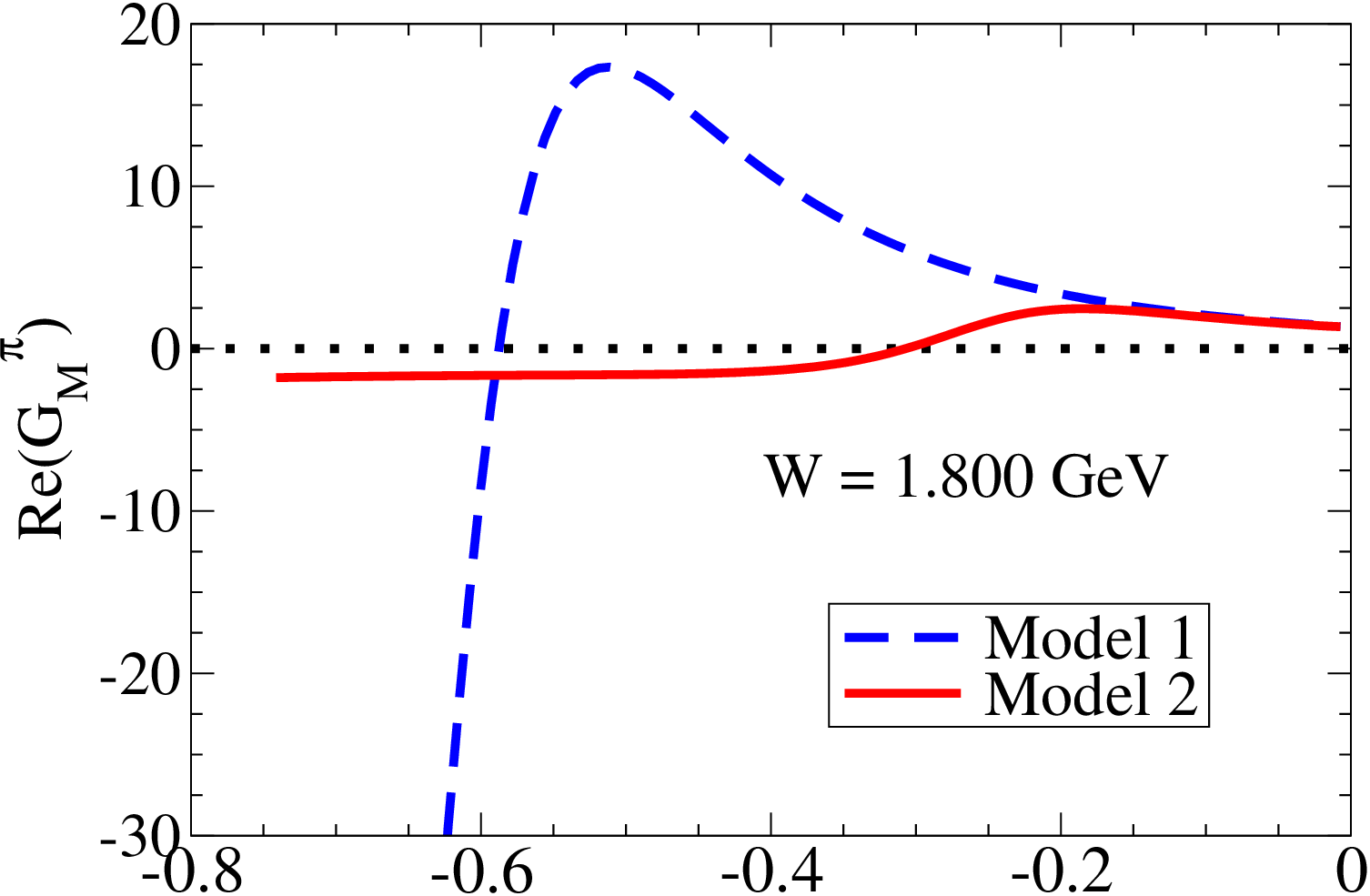} \hspace{.8cm}
\includegraphics[width=2.8in]{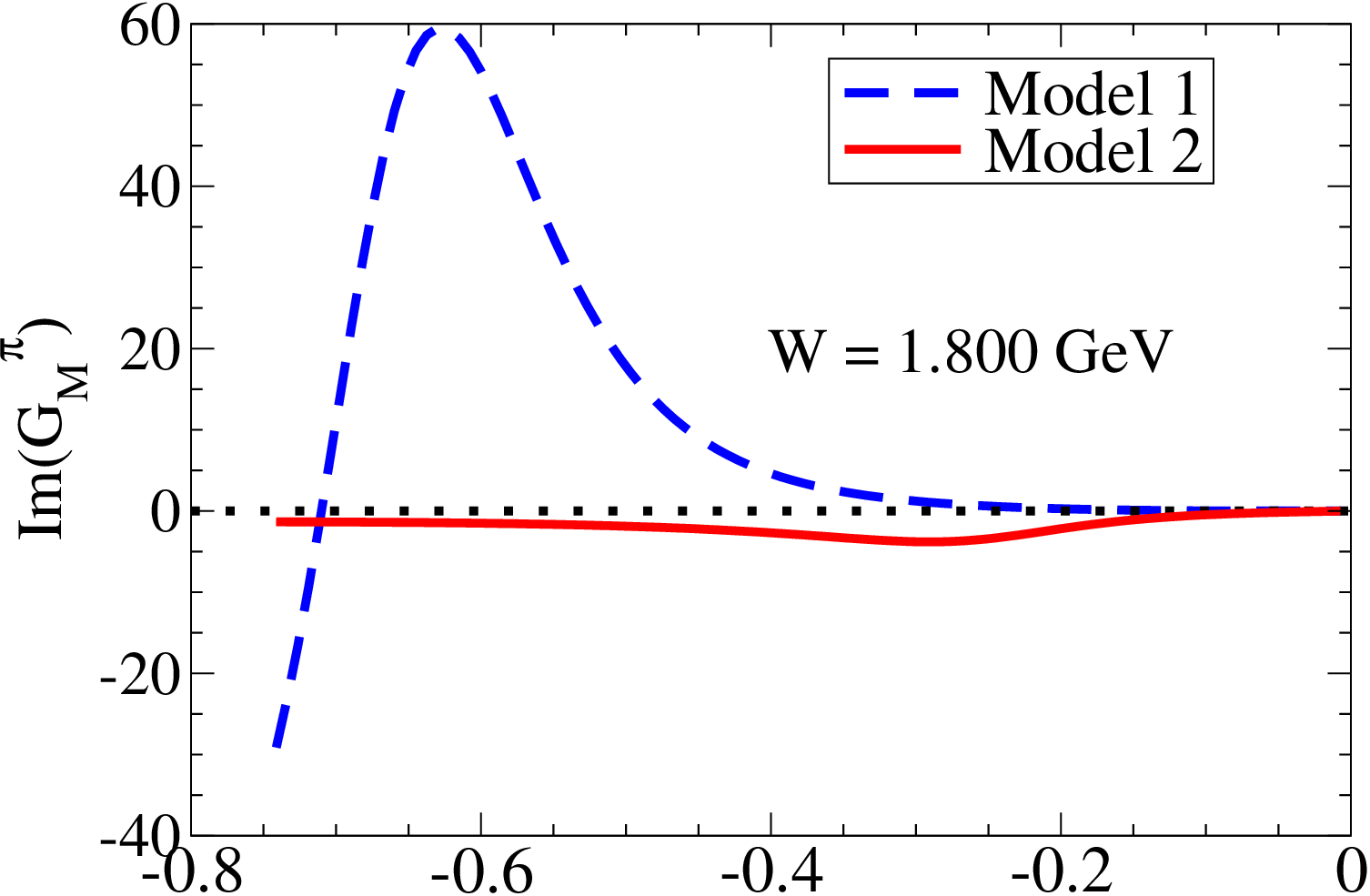}
}}
\centerline{
\vspace{.5cm} }
\centerline{
\mbox{
\includegraphics[width=2.8in]{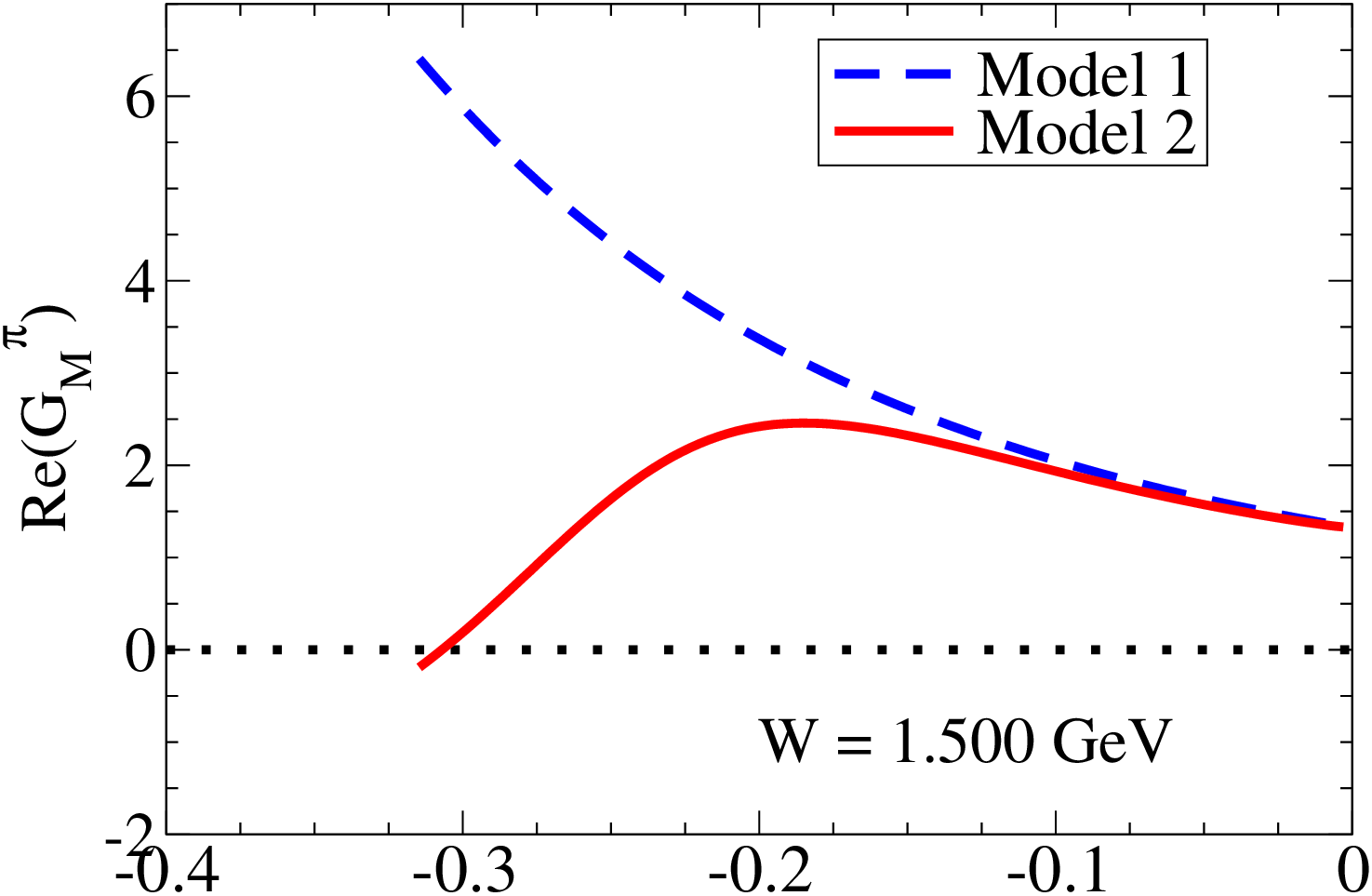} \hspace{.8cm}
\includegraphics[width=2.8in]{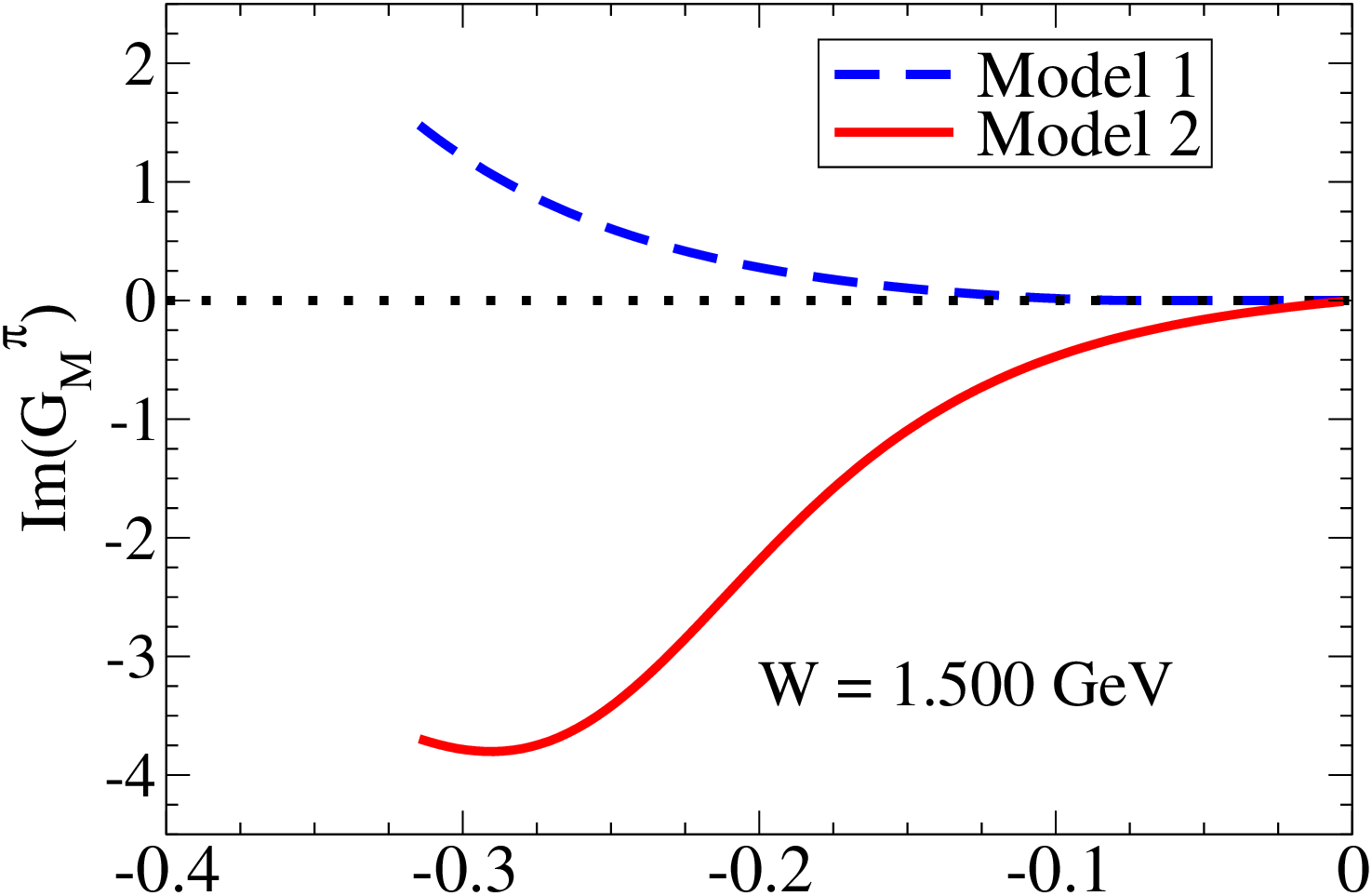}
}}
\centerline{
\vspace{.5cm} }
\centerline{
\mbox{
\includegraphics[width=2.8in]{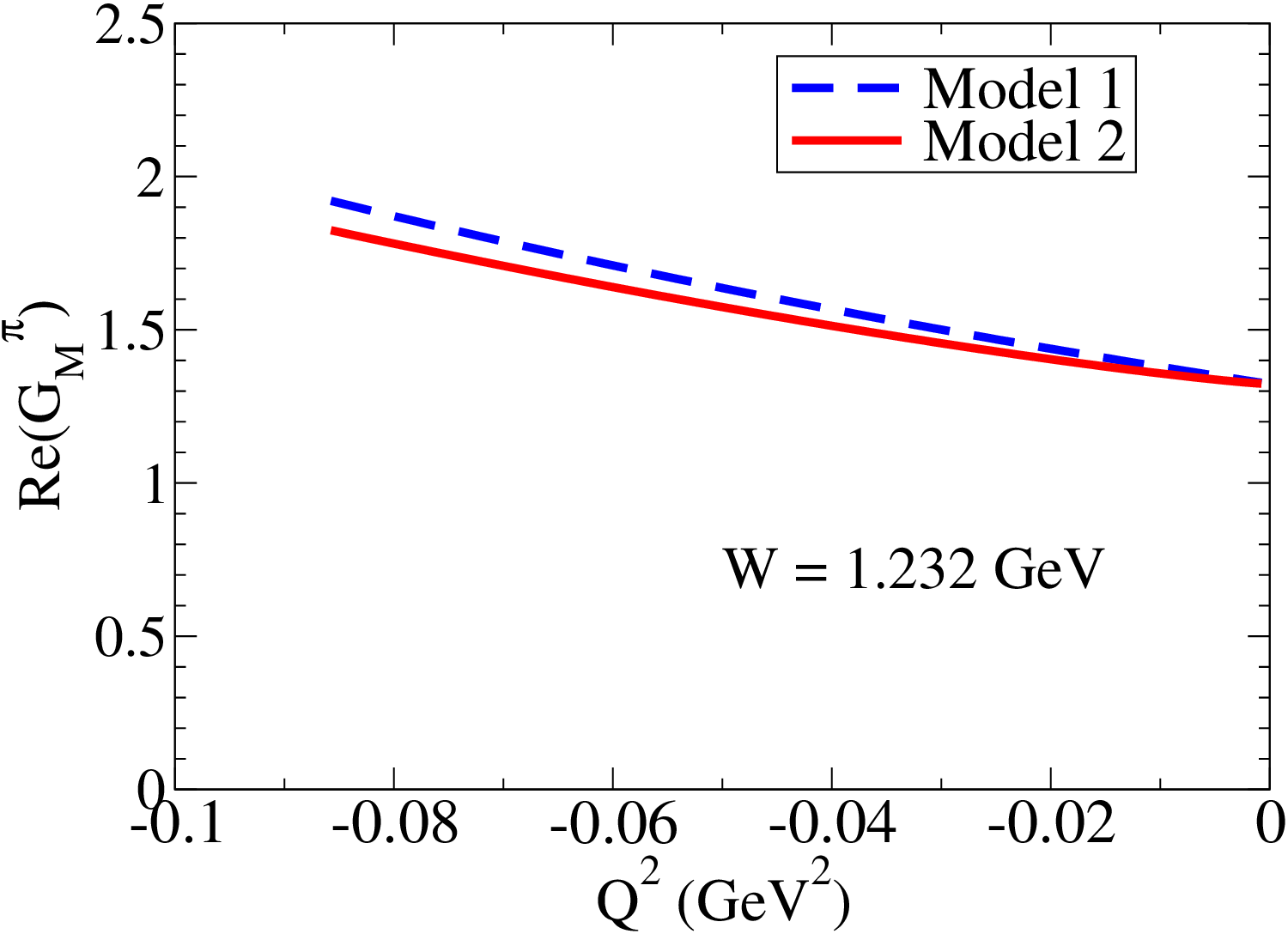} \hspace{.8cm}
\includegraphics[width=2.9in]{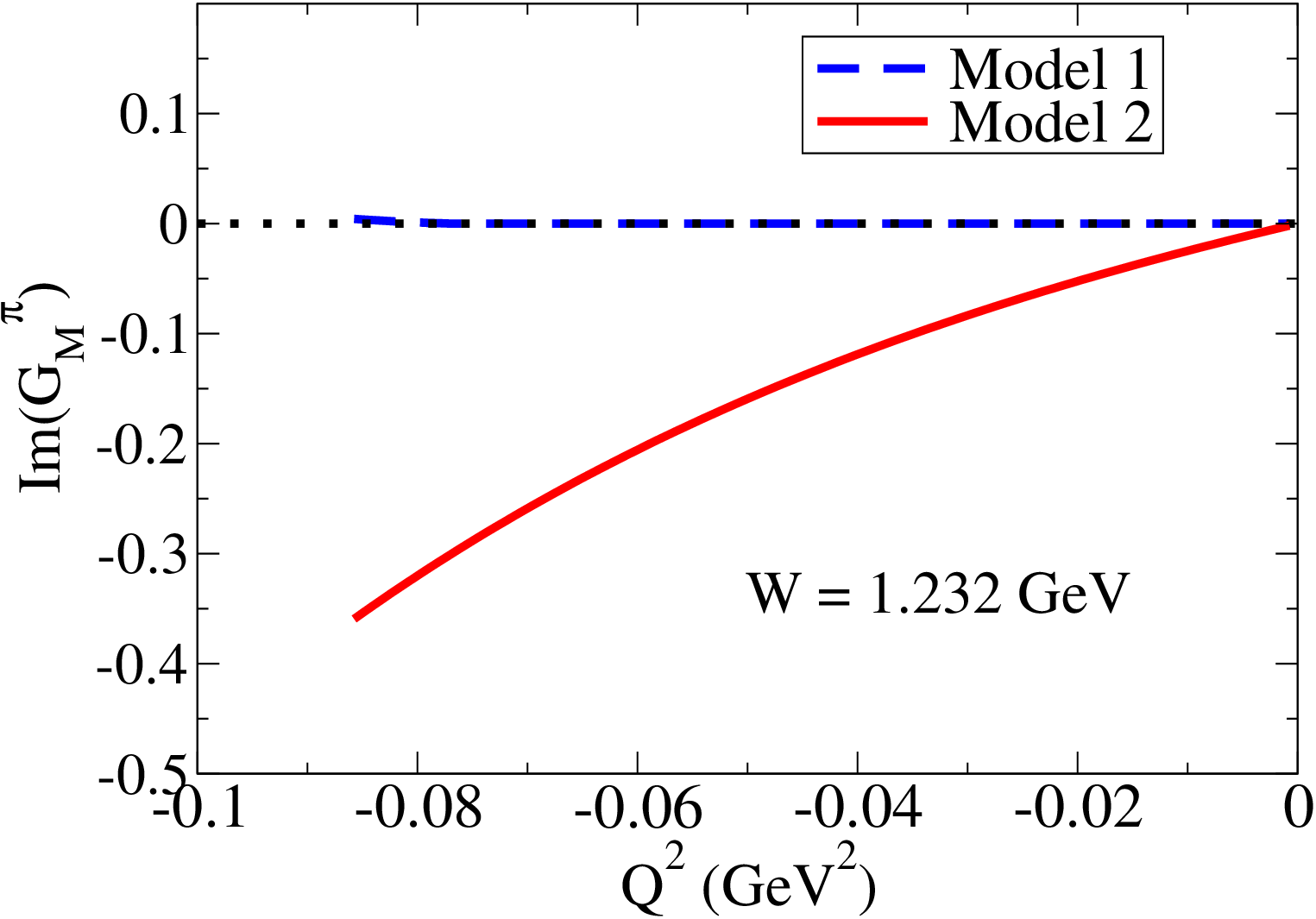}
}}
\caption{\footnotesize{
Pion cloud contributions ($G_M^\pi$) for the 
$G_M^\ast$ form factor (real and imaginary parts).
The results from models 1 and 2 are shown.
}}
\label{figGMpi}
\end{figure*}

We turn now to the term $G_M^\pi$ which gives the
pion cloud contribution, and where the two models 
differ in the timelike region.
The results are presented in Fig.~\ref{figGMpi}
for the same values of $W$ as before. 
Comparing the two models,  
we can say that both 
have similar results for 
the real part of $G_M^\pi$ in the 
region \mbox{$-0.2$ GeV$^2 <Q^2$}, but 
differ significantly for smaller values of $Q^2$ (larger values of $q^2$).
That is the consequence of the double pole $q^2=\Lambda_D^2$
in the pion cloud formula for model 1.
In model 2 there is no such contribution from the pion cloud
and the values for real and imaginary parts are more moderate.
Note that the strong peak for $W=1.800$ GeV 
at $Q^2 \simeq -0.65$  GeV$^2$,
for model 1 is a consequence of Eq.~(\ref{eqGD2}), and differs from model 2
by  an order of magnitude.

We look now to the imaginary part of $G_M^\pi$.
Our first observation goes to 
the imaginary part of $G_M^\pi$ in the region near $Q^2=0$.
In model 1 it is identically
zero for $-0.076 \le Q^2 \le 0$  
[because $\Gamma_\rho = 0$ from Eq.~(\ref{eqGammaR})],
but in model 2 it is  different from zero, although small,
(this is a consequence of the approximation 
considered in function $F_\rho$ discussed previously).
The second observation is that 
the significant difference 
between models 1 and 2 is the 
sign of the imaginary part of $G_M^\pi$:  model 1 
gives positive contributions, while
model 2 gives negative contributions. 
This model is motivated by  $\chi$PT, 
satisfies chiral constraints for the $\rho$ propagator, which 
has a non-analytical pole near 
$Q^2 \simeq - 0.3$ GeV$^2$ 
present in $F_\rho$ and with its origin in the pion loop contributions.

With this detailed analysis we come to understand  
the large difference between 
the two models shown in Fig.~\ref{figGM2}. 
The figure provides a strong indication that 
model 1 is not a reasonable model: the results from this model
are strongly dominated by the pole $q^2=\Lambda_D^2=0.71$ GeV$^2$
which induce extremely  large (and probably unphysical) 
contributions for increasingly large $W$ values.
On the other hand, model 2 contains 
the input from $\chi$PT, and has therefore a
more solid basis.
The results are very sensitive to this input and clearly exclude model 1.

\begin{figure}[t]
\vspace{.3cm}
\centerline{
\mbox{
\includegraphics[width=2.8in]{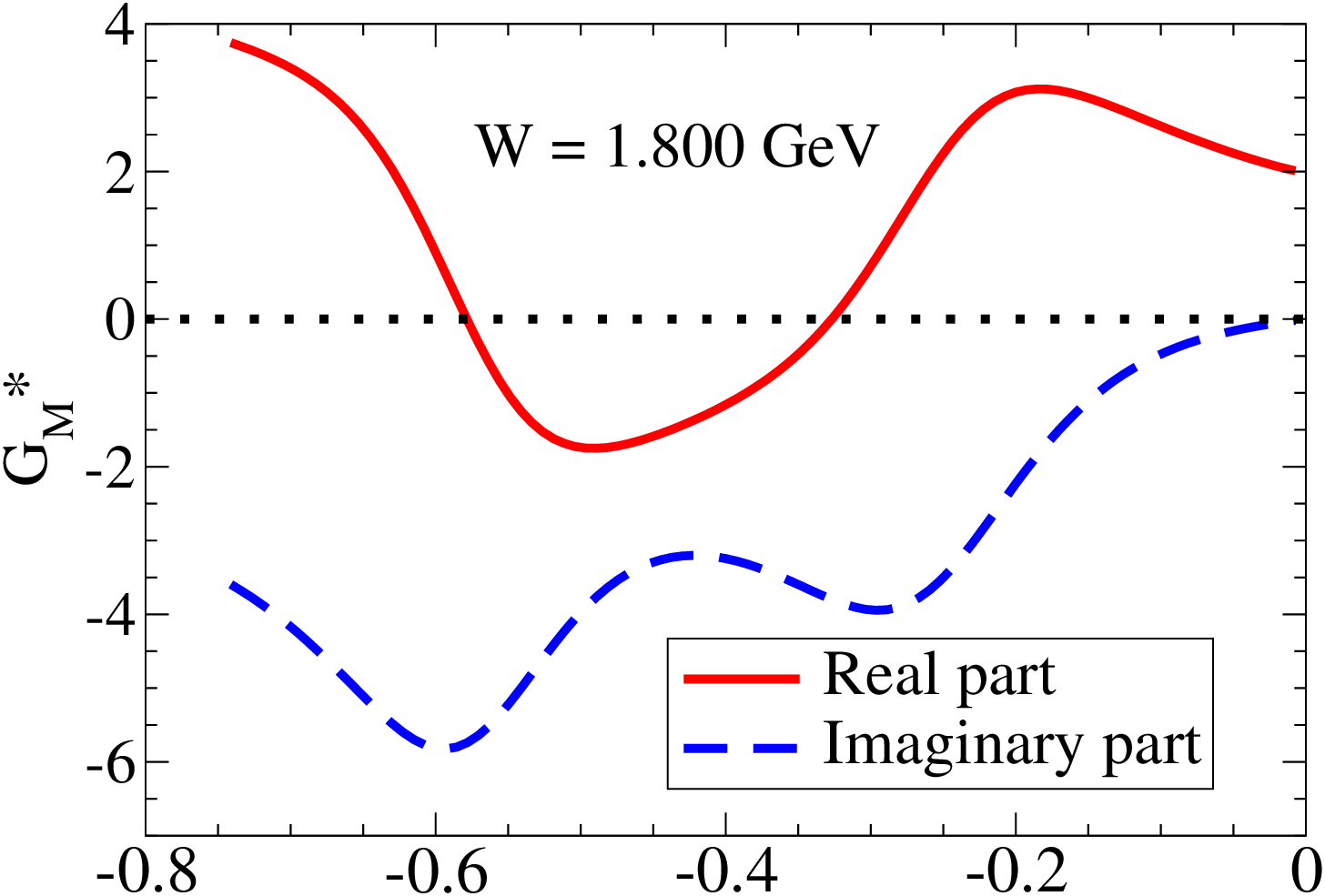}
}}
\centerline{
\vspace{.5cm} }
\centerline{
\mbox{
\includegraphics[width=2.8in]{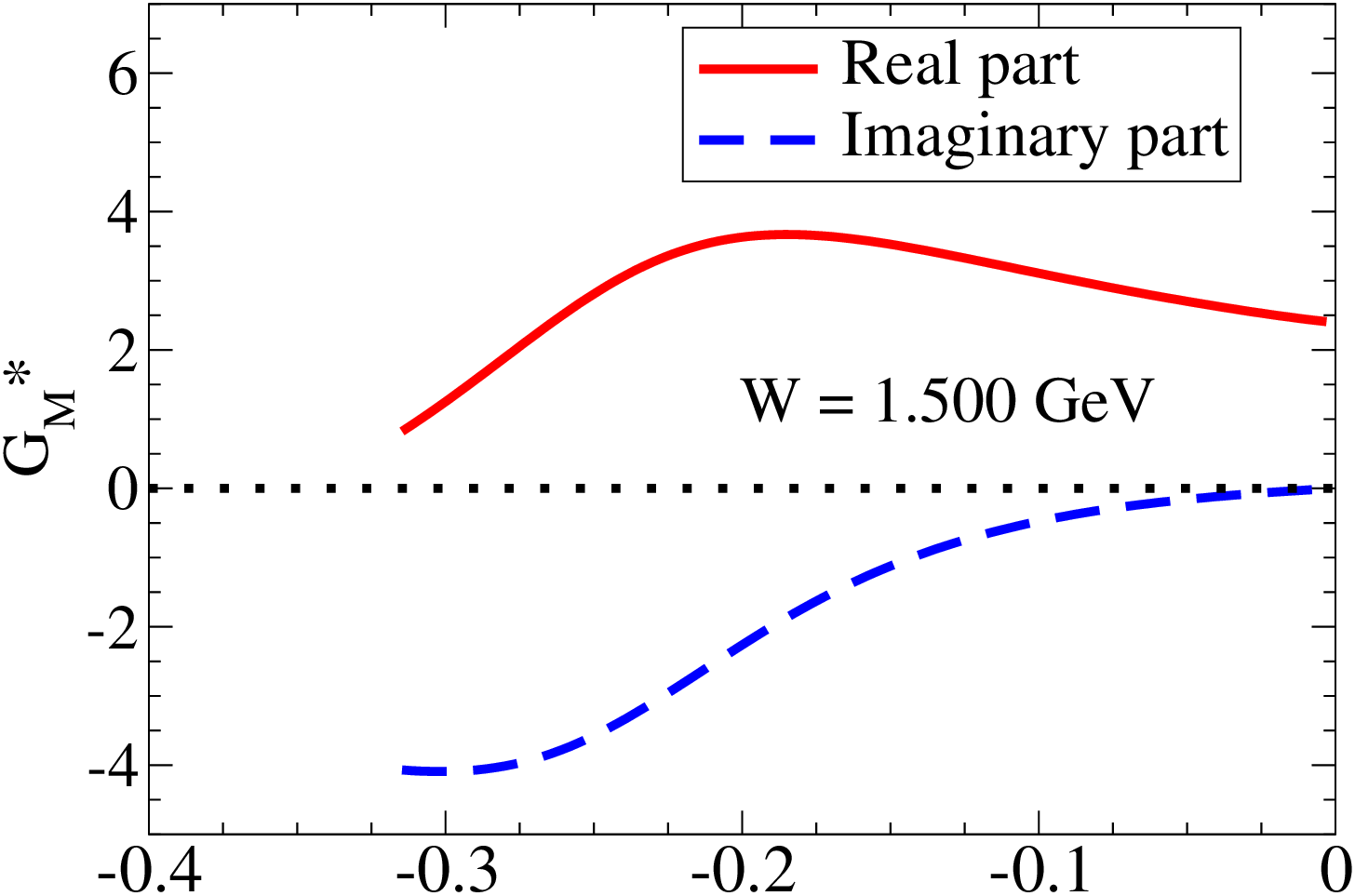}
}}
\centerline{
\vspace{.45cm} }
\centerline{
\mbox{
\includegraphics[width=2.8in]{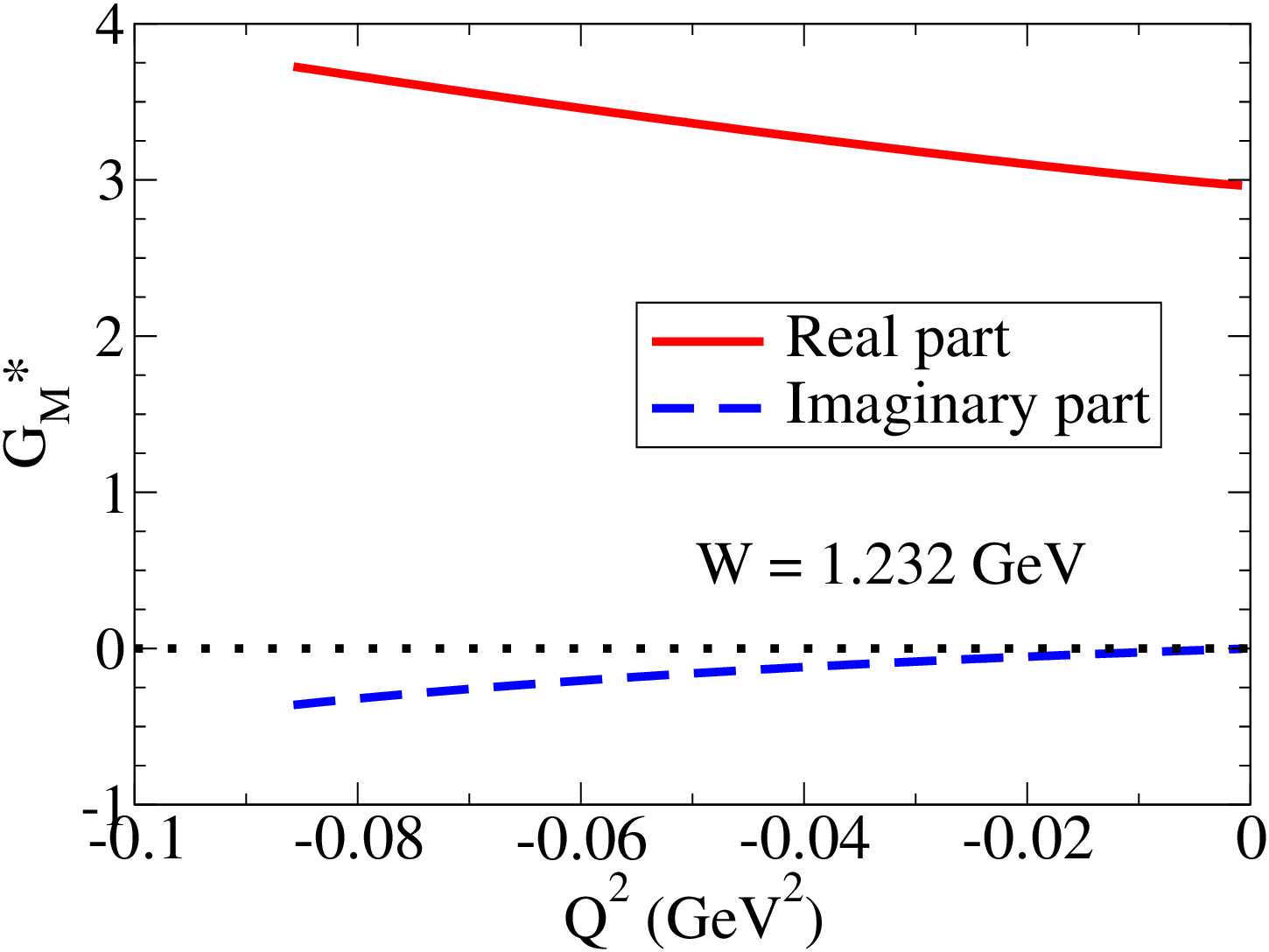}
}}
\caption{\footnotesize{
Real and imaginary components of the form factor 
$G_M^\ast$  for model 2.
}}
\label{figGMmod2}
\end{figure}

From the previous discussion we favor
the results from model 2.
The final results (bare quark core plus pion cloud) 
for both the real and imaginary part of
the form factor $G_M^\ast$ 
are presented in Fig.~\ref{figGMmod2}. 
One realizes that the 
real part dominates over the imaginary part 
for $Q^2 > -0.15$ GeV$^2$. 
We can say
that the dominant contributions
to the imaginary part are the poles 
from $G_M^\pi$ (induced by $F_\rho$) around $Q^2 =- 0.3$ GeV$^2$,
and the $\rho$-terms in the quark current from the bare contribution.
The effect of those poles is particularly 
evident in the curve for $W=1.800$ GeV.

The main and general conclusion from Figs.~\ref{figGMB}, 
\ref{figGMpi} and \ref{figGMmod2}   
is that the final  structure for  $G_M^\ast$ emerges from 
a combination of effects,
namely from the VMD model poles and also 
the pion cloud effects.
The two processes interfere crucially and determine the structure 
of the final amplitude.

Later, we will discuss the applicability 
of the model for $W> 1.8$ GeV, 
the effect of the remaining poles,
and the impact in the observables in consideration.
We emphasize that any extension of 
the $\Delta$ form factors to the timelike region 
has to rely on models and cannot be directly estimated 
from experimental data only. 
It is then important to compare 
our model with models with a similar content,
such as the  two-component quark model of Ref.~\cite{Frohlich10}, 
also defined in the timelike region.  
In this last model the contribution from the coupling 
to the quark core (valence contribution) is
0.3\% near $Q^2=0$ (99.7\% of pion cloud),
while in our model one has 55.9\% (44.1\% of pion cloud).
This significant difference between the contributions 
of the quark core is due to a different, 
and  somewhat arbitrary, classification of the two effects.
In the model of Ref.~\cite{Frohlich10,Iachello04} the term from the 
pion cloud is also classified as an effective part of the
VMD mechanism, since it is proportional to the 
function $F_\rho$ for the $\rho$ propagator.
Therefore, in that model
the VMD mechanism/pion cloud term
is the only relevant effect~\cite{Frohlich10,Iachello04}.
In our formalism, the coupling with the quarks 
is calibrated directly by a VMD parametrization 
and although it gives the dominant contribution, 
it is not the only one to affect the results. 
Our model has the advantage of having been tested 
successfully by the lattice QCD simulations  
(in a regime where the pion cloud is small),
and of agreeing with the EBAC data analysis for the bare
quark core  contributions to the pion photoproduction data~\cite{LatticeD}.
These tests
suggest that our estimation 
of the quark core structure is under control, 
since the model is largely constrained in a variety of 
kinematic domains.
Another important point is that our model allows a direct 
physical interpretation of the parameters involved, in terms of the range
of the baryon wave functions.

\subsection{Results for $\Gamma_{\gamma N}(W)$ 
and $\Gamma_{e^+ e^- N}(W)$ }

We will discuss now the partial widths $\Gamma_{\gamma N}(W)$ 
and $\Gamma_{e^+ e^- N}(W)$.
We will also show $\Gamma_{\pi N}(W)$, given by Eq.~(\ref{eqGammaPiN}),
together with the calculation of $g_\Delta(W)$, defined by Eq.~(\ref{eqG}).

\begin{figure}[t]
\vspace{.3cm}
\centerline{
\mbox{
\includegraphics[width=3.1in]{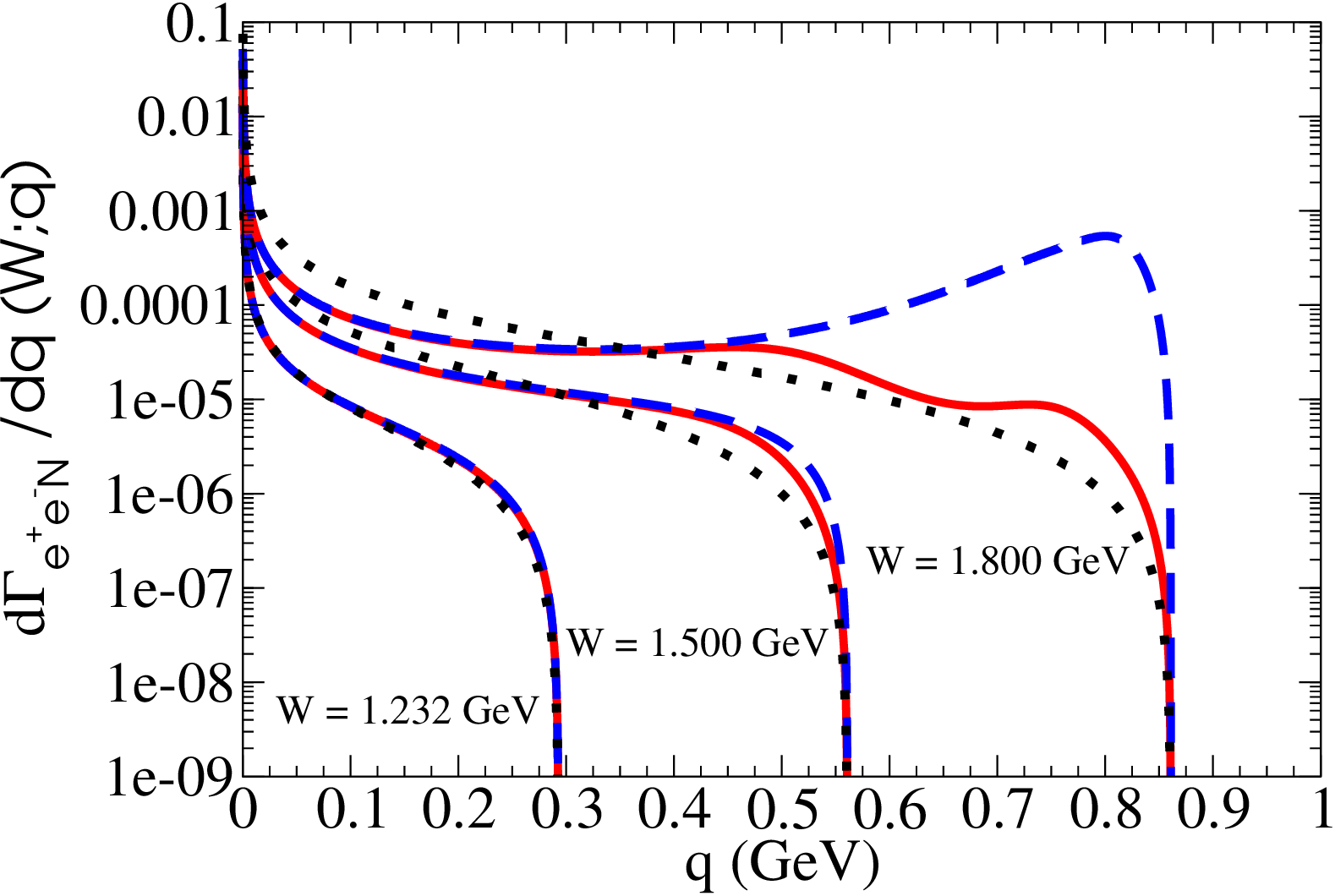}
}}
\caption{\footnotesize{
The distribution $\frac{d \Gamma_{e^+ e^- N}}{dq}(W;q)$
as defined in Eqs.~(\ref{eqGamma1}) and (\ref{eqGamma2})
for 3 different energies $W$.
The solid line is the result of model 2;
the dashed line is the result of model 1 (naive model).
The dotted line is the result obtained by using a constant form factor. 
The function has no dimensions.
}}
\label{figDGamma}
\end{figure}

We start by showing in Fig.~\ref{figDGamma} the function  
$\frac{d \Gamma_{e^+ e^- N}}{dq}(W;q)$ 
for the cases $W=1.232, 1.500$ and 1.800 GeV.
This figure includes the results from model 1 (dashed line), 
model 2 (solid line) and also the result 
of a calculation  where  the form factor $G_M^\ast$  is taken as constant, 
defined by the value of $G_M^\ast(W,0)$
at the pole $W=M_\Delta$ (dotted line), 
given by the experimental value
[$G_M^\ast(Q^2) \equiv G_M^\ast(M_\Delta,0) \simeq 3.0$].
This last case was also considered in Ref.~\cite{Frohlich10}
and it is useful as a reference for the $q^2$ dependence of our results.
The figure illustrates that, in line with the results 
in the previous subsection, for model 1 
$\Gamma_{e^+ e^- N}^\prime (q,W)$ is enhanced for large $q$
and large $W$ values (see result for $W=1.800$ GeV).

\begin{figure}[t]
\vspace{.4cm}
\centerline{
\mbox{
\includegraphics[width=3.0in]{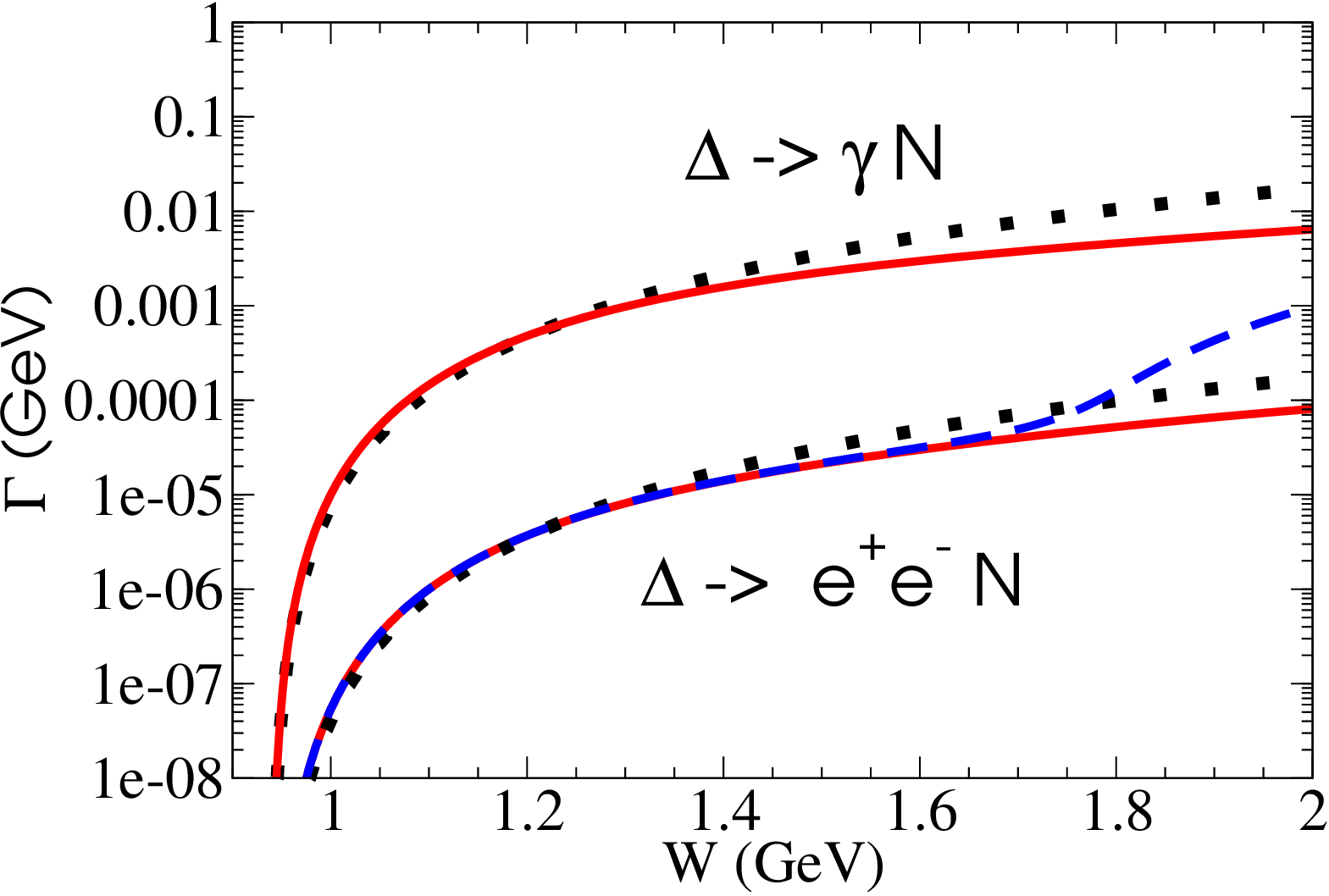}
}}
\caption{\footnotesize{
Partial widths as function of $W$ for different models.
The constant form factor model is 
represented by the doted line;
the model 1 by the dashed line
and the model 2 by the solid line.
Note that models 1 and 2 have the same result 
for the real decay  $\Delta \to \gamma N$, 
because the models are identical at $Q^2=0$.}}
\label{figGamma1}
\end{figure}

To determine the di-lepton production width, $\Gamma_{e^+ e^- N}(W)$, one has to 
integrate Eq.~(\ref{eqGamma2}) 
using Eq.~(\ref{eqInt1}).
This is equivalent to calculate the integral 
of the functions represented in  Fig.~\ref{figDGamma}
for each value of $W$ in the interval $[2 m_e, W-M]$.
Therefore $\Gamma_{e^+ e^- N}(W)=0$ when $W< M + 2 m_e$.
The calculation of the function $\Gamma_{\gamma N}(W)$ proceeds
through Eq.~(\ref{eqGamma0}).
The results obtained for the two widths within 
the three models discussed before, 
are in Fig.~\ref{figGamma1}.

\begin{figure}[t]
\vspace{.3cm}
\centerline{
\mbox{
\includegraphics[width=3.0in]{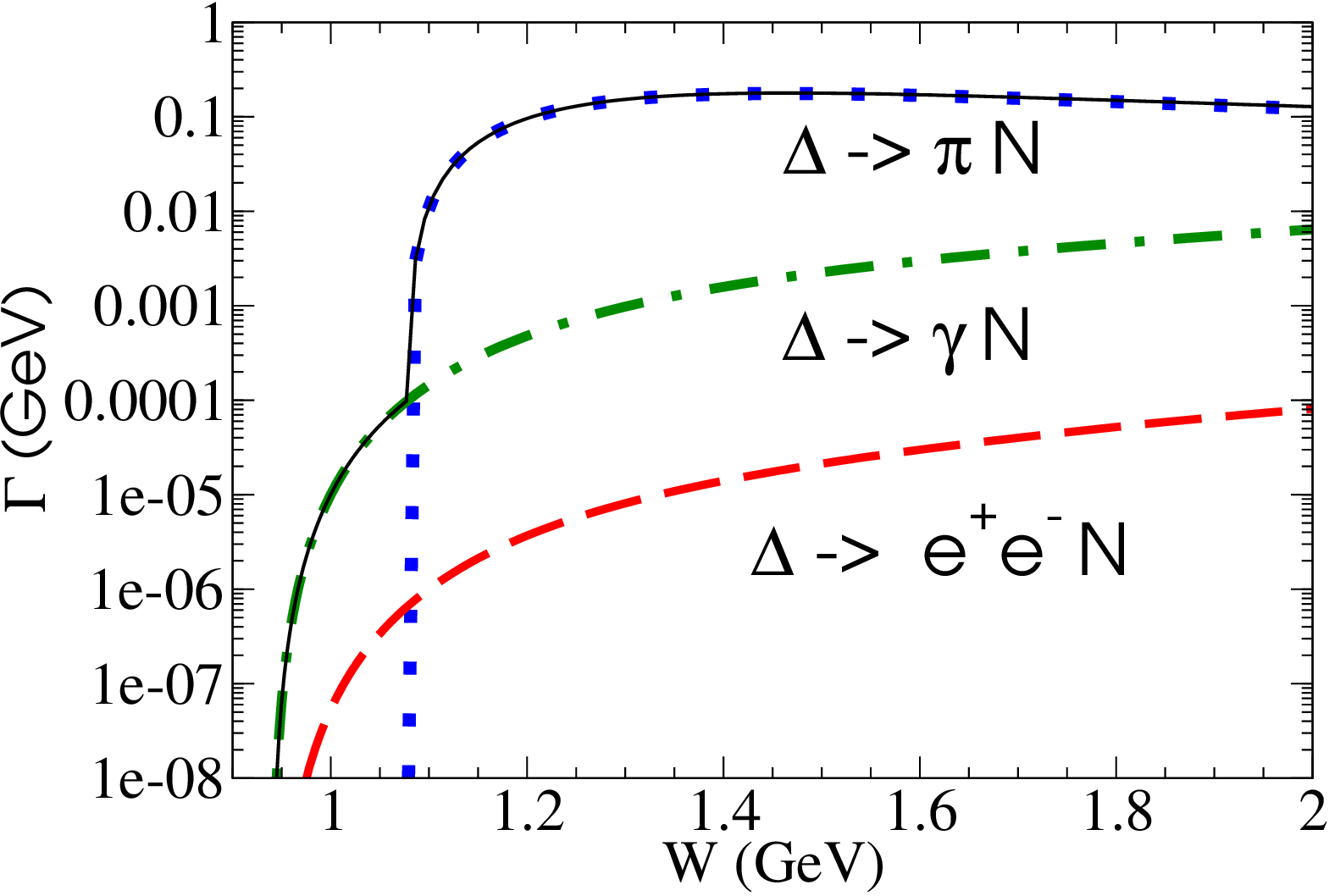}
}}
\caption{\footnotesize{
Partial widths as function of $W$ for model 2. 
The total is represented by the thin solid line.}}
\label{figGamma2}
\end{figure}

Finally,  $\Gamma_{\pi N}(W)$ is estimated using Eq.~(\ref{eqGammaPiN})
and the function
\ba
q_\pi (W)=
\frac{\sqrt{\left[(W+M)^2-m_\pi^2 \right] \left[(W-M)^2-m_\pi^2 \right]}}{2W},
\ea
defined for $W \ge M+ m_\pi$ and $q_\pi(W)=0$ otherwise. 
$\Gamma_{\pi N}(W)$ is then
a positive function for $W> M + m_\pi$.
In Fig.~\ref{figGamma2} we present the three partial 
widths obtained with model 2, the one that 
we favor for the reasons explained in the previous subsection.

We turn now to the 
$\Delta$ mass distribution function $g_\Delta(W)$
defined by Eq.~(\ref{eqG}).
As the channel $\Delta \to \pi N$ 
is largely dominant, $\Gamma_{tot}(W) \simeq \Gamma_{\pi N}(W)$
and the normalization of $g_\Delta(W)$ can be done in that 
approximation. 
Considering 
$\Gamma_{\pi N} (M_\Delta) \simeq \Gamma_{tot}(M_\Delta) \simeq 
\Gamma_\Delta^{exp}$, with 
the experimental result $\Gamma_\Delta^{exp} \simeq 0.118$ GeV \cite{PDG},
one has $A=0.7199$.

\begin{figure}[t]
\vspace{.3cm}
\centerline{
\mbox{
\includegraphics[width=3.0in]{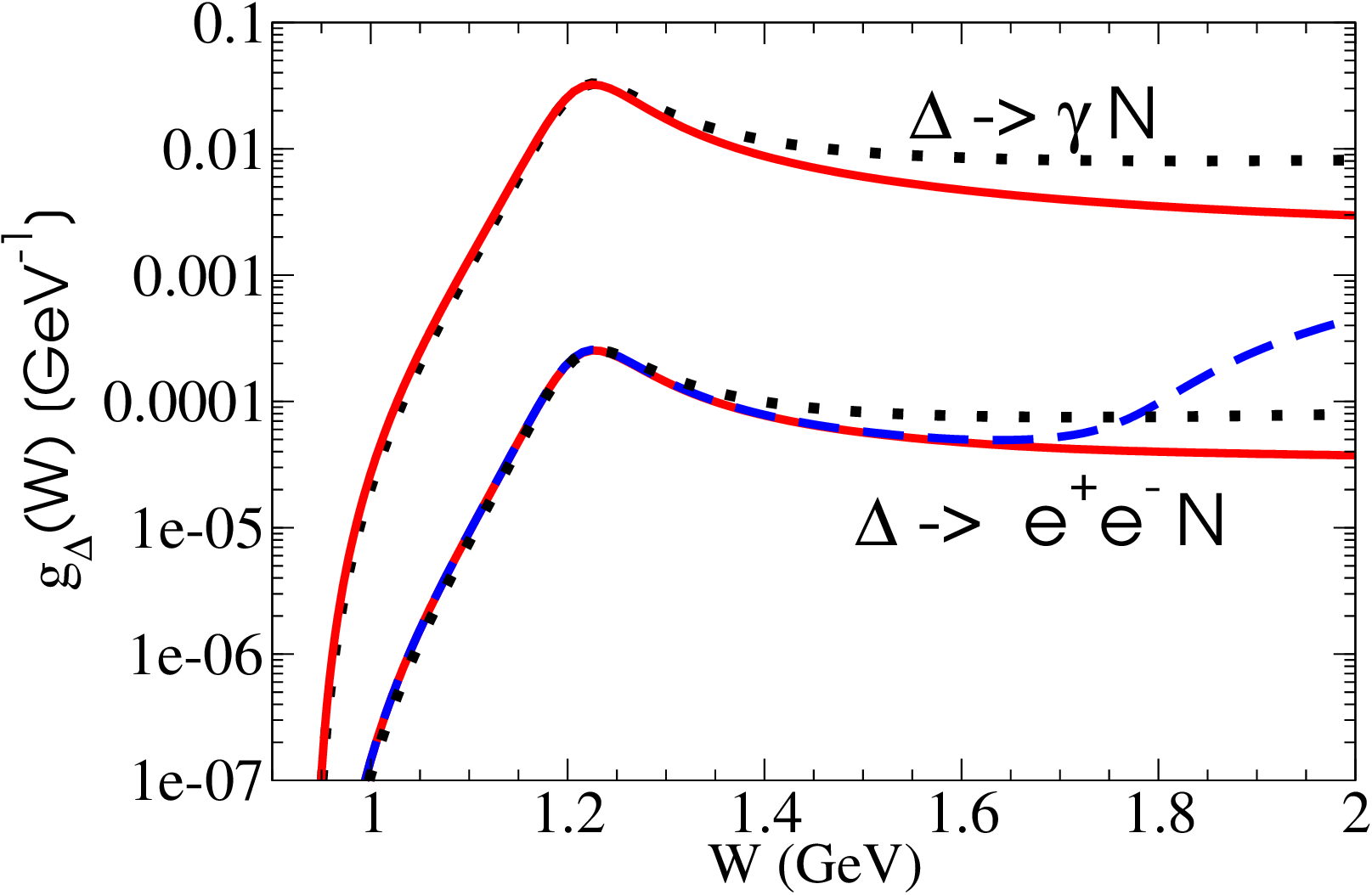}
}}
\caption{\footnotesize{
Partial contributions 
of the channels $\Delta \to \gamma N$ and  $\Delta \to e^+ e^- N$
(in GeV$^{-1}$ units).
The constant form factor model is 
represented by the doted line;
the model 1 by the dashed line
and the model 2 by the solid line.
Note that models 1 and 2 have the same result 
for the real decay  $\Delta \to \gamma N$, 
because the models are identical at $Q^2=0$.
}}
\label{figGD1}
\end{figure}

\begin{figure}[t]
\vspace{.3cm}
\centerline{
\mbox{
\includegraphics[width=3.0in]{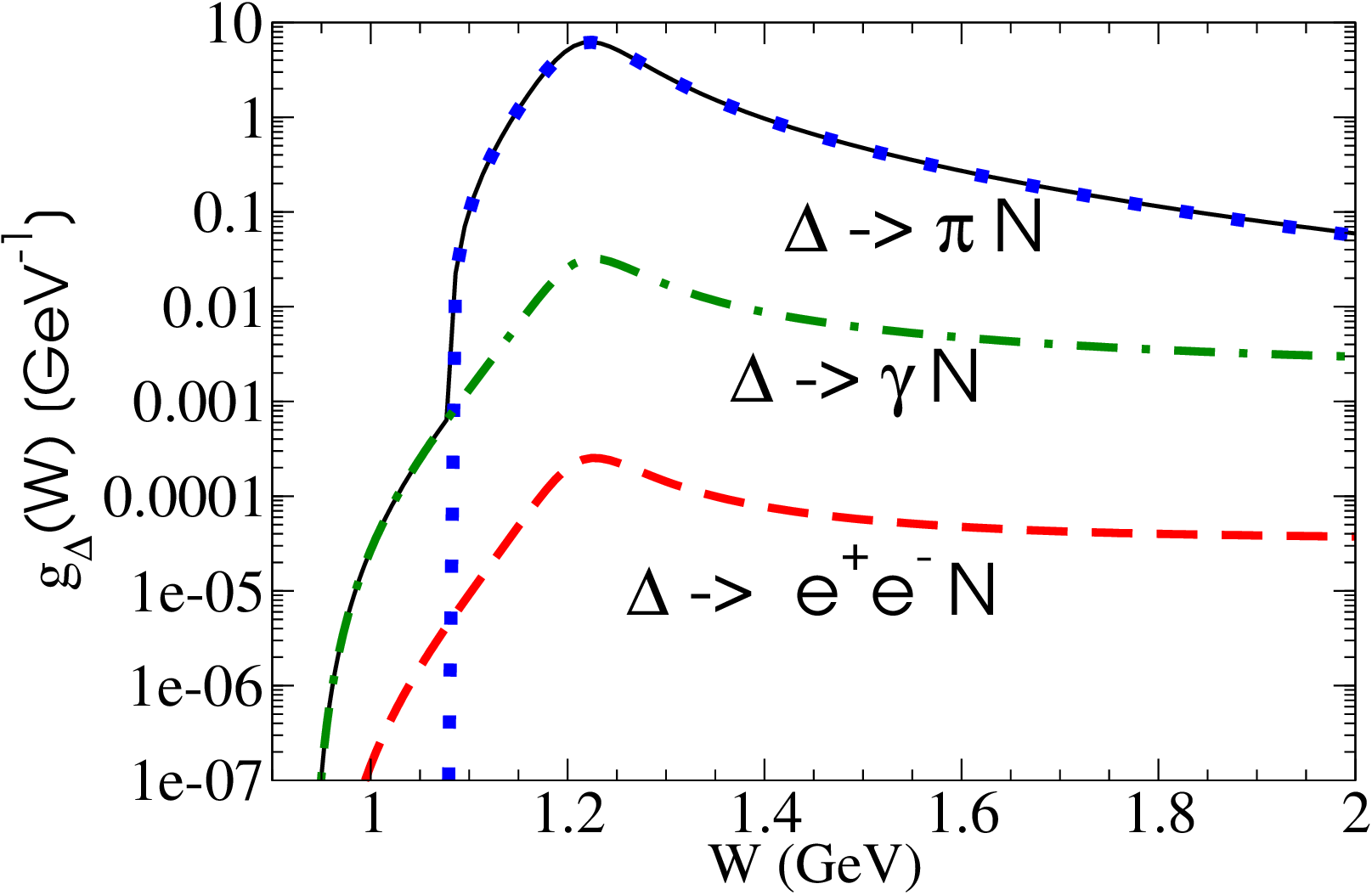}
}}
\caption{\footnotesize{
Function $g_\Delta(W)$ (in GeV$^{-1}$ units)
and the partial contributions
of the channels $\Delta \to \gamma N, \Delta \to e^+ e^- N$
as function of $W$ for the model 2.
The thin solid line is the total result ($g_\Delta(W)$) 
and the sum off all decay channels 
($\Delta \to \pi N, \Delta \to \gamma N$ and 
$\Delta \to e^+ e^- N$).
}}
\label{figGD2}
\end{figure}

The results for the partial contributions to
$g_\Delta(W)$ are given by
 \ba
& &
g_{\Delta \to \gamma N}(W)=
\frac{\Gamma_{\gamma N}(W)}{\Gamma_{tot}(W)} g_\Delta(W) 
\nonumber \\
& &
g_{\Delta \to e^+ e^- N}(W)=
\frac{\Gamma_{e^+ e^- N}(W)}{\Gamma_{tot}(W)} g_\Delta(W), 
\ea
and are shown in Fig.~\ref{figGD1} for 
the constant form factor model (dotted line),
model 1 (dashed line) and model 2 (solid line).
The total results for $g_\Delta(W)$ are 
shown in Fig.~\ref{figGD2}, for the model 2.

We restricted our results to the region $W\le 2$ GeV.
Above that region one has to take into account 
the additional pole structure 
of the form factor $G_M^\ast$ that appears for
large $q^2$ values.
Another reason for not having our model applied to
larger $W$ values is that, for $W > 1.5$ GeV, the reactions 
are expected to be dominated by resonances as the 
$N^\ast(1440)$, $N^\ast(1535)$, $\Delta(1600)$ among others,
instead of the $\Delta(1232)$ alone.  

As for the first problem, one has to find a way 
deal with those large $q^2$ singularities. 
From Eq.~(\ref{eqGMpi2}) for the pion cloud component, already,
the pole at $q^2=\Lambda_{\pi}^2 \simeq 1.53$ GeV$^2$, makes the form factor
diverge for $W>M+\Lambda_{\pi} \simeq 2.17$ GeV, since $q^2 \le (W-M)^2$.
Also, from the valence quark component one has a 
singularity $q^2= M_h^2 \simeq 3.53$ GeV$^2$.
To remove those singularities
we may introduce an effective width $\Gamma_X$,
in analogy with Eq.~(\ref{eqVMD})
or (\ref{eqGD2}), for single or double poles, 
with a constant width $\Gamma_X^0$.
This procedure adds a new parameter to the model.
We started by verifying that
a large width $\Gamma_X^0$ (for instance $\Gamma_X^0 \simeq 1$ GeV)
will affect the results  for $W <2$ GeV, while
a very small width 
will induce significant oscillations 
 in the functions 
$g_{\Delta \to \gamma N}(W)$ and $g_{\Delta \to e^+ e^- N}(W)$
(enhancement of $\Gamma^\prime_{e^+ e^- N}(q,W)$ near the poles).
By taking $\Gamma_X^0 = 2 \Gamma_\rho^0$
the results obtained for $W < 2$ GeV 
are almost unchanged, and the results for 
$W >2 $ GeV are also smooth functions of $W$.
With this choice
the smooth dependence on $W$ of the functions
 $g_{\Delta \to \gamma N}(W)$ and $g_{\Delta \to e^+ e^- N}(W)$,
obtained with model 2 is maintained for higher $W$ values.
We stress, nevertheless, that the application 
of our model to large $W$ values is to be viewed
only as an extreme test to its limits.

\section{Conclusions} 
\label{secConclusions}

In this work we have presented a covariant approach to describe the
the Dalitz decays $\Delta \to \gamma N$, $\Delta \to \gamma^\ast N$,
and we have calculated the 
$\Delta$ mass distribution function $g_\Delta(W)$ within that approach.
Our framework can be used to simulate the
$NN \to e^+ e^- NN$ reaction 
at  moderate beam kinetic energies ($1-2$ GeV).
The code used in this work can be supplied 
under request.\footnote{Send an email to 
{\bf gilberto.ramalho@cftp.ist.utl.pt}.}

Our calculations are based on an unified description
of the   $\gamma^\ast N \to \Delta$ reaction,  in both the 
spacelike and timelike regimes.
We start with a model tested previously 
in spacelike physical and lattice QCD simulation data \cite{LatticeD,NDeltaD},
and generalize it to the timelike regime.
In this formalism the electromagnetic interaction 
can be decomposed into two mechanisms:
the direct photon coupling with the quarks 
and the interaction with the pion cloud.

For the first mechanism we extended the quark current and 
wave functions, obtained for $q^2<0$, to the region $q^2> 0$ 
without additional changes, except for a non-zero width 
of the effective vector mesons included in the VMD parametrization 
of our quark electromagnetic current. For the pion cloud contribution 
we probed two different parametrizations:
a naive generalization of our spacelike model
and a more elaborated model based on $\chi$PT.
Although the two models behave very similarly in the spacelike region, 
they differ substantially  in the timelike region.

The results of the models for the form factor 
$G_M^\ast$, as a function of $q$ and $W$, are used 
to calculate the partial widths $\Gamma_{\gamma N}(W)$, 
$\Gamma_{\Delta \to e^+ e^- N}(W)$ and the 
partial mass distributions functions $g_{\Delta \to \gamma N}(W)$,
$g_{\Delta \to e^+ e^- N}(W)$.

A first important conclusion of this work is that the $q^2$ 
dependence of the form factor 
has an impact in the final results, and has therefore to be under control:
the results for the 
$\Delta$ mass distribution functions, 
where the form factor is taken with its full
$q^2$ dependence, can differ by a factor of 4 from the results
obtained with the constant form factor model. 

We verified also that the results are
sensitive to the analytical extension
of the pion cloud parametrization to the timelike region.
Model 1 (naive model) generates unreasonably large 
contributions 
for moderate squared momentum $q^2$,
for larger values of $W$, 
as consequence of a spurious (not physically motivated) 
pole $q^2= \Lambda_D^2$ in the pion cloud contribution. 
Model 2 is motivated by $\chi$PT and 
includes the function $F_\rho$ for the $\rho$ propagator with a 
non-algebraic pole near $q^2 \simeq 0.3$ GeV$^2$ 
originated by pion loop corrections.
This model gives smooth contributions to
the partial 
$\Delta$ mass distribution functions
$g_{\Delta \to \gamma N}(W)$ and $g_{\Delta \to e^+ e^- N}(W)$
that vary slowly with $W$ for large $W$.

A second important conclusion is then
that our calculations 
support the need to have under control  
the effect of the pion cloud contributions.  
Our framework is suitable for this because its
bare quark core component is constrained by 
experimental data 
and lattice QCD simulation data.
In addition, the fact that the pion cloud content of model 2 
is consistent with $\chi$PT
makes model 2 reliable, at least in its domain of validity.
The results of model 2
 give moderated 
contributions for $|G_M^\ast|^2$ 
(since it has no spurious pole at $q^2=\Lambda_D^2$), which are determined
by the combined
effect of two important features:
the pion cloud term structure near
$q^2=0.3$ GeV$^2$,  and the quark core contributions 
from the $\rho-$pole included 
in the VMD structure of the quark current, near $q^2=0.6$ GeV$^2$.

We may say that while spacelike data 
does not constrain models sufficiently well enough, timelike data for 
$\sfrac{d \Gamma_{e^+ e^- N}}{d q}$
for different values of $W$ (see Fig.~\ref{figDGamma}) 
are important and necessary to select between models 
in a decisive way.
In the $\Delta$ case discussed in this work this is  
specially true for the pion cloud effects.
But the timelike data can also be useful
to calibrate the widths and high mass poles 
of the VMD parametrization of the current 
needed in valence quark component,
particularly for resonances heavier 
that the $\Delta(1232)$.

For high values of $W$ the assumption that
the $\Delta(1232)$ resonance is the only state playing 
a role in the reactions
becomes questionable.
In the regime $W> 1.5$ GeV
other resonances can be relevant, 
as the spin 1/2 resonances $N^\ast(1440)$ and $N^\ast(1535)$. 
Once those states 
are calibrated for the $Q^2=0-2$ GeV$^2$ region, one can extend 
the models from Refs.~\cite{Roper,S11} 
to the timelike region too.
It is also
expected that the spin  3/2 channels as the $\Delta(1600)$ state are important
for large $W$.
This defines a study of high interest, 
since the constraints in the spacelike region 
are very scarce, and the available data at 
the photon point suggest a strong contribution 
from the pion cloud \cite{Delta1600}.

Future applications of our formalism 
can include the study of the $\gamma^\ast N \to N$ reaction 
where the final nucleon has an arbitrary mass $W$,
also in the timelike region.
Since the quark current was already defined 
in the timelike region and we have already  
a model for the nucleon system \cite{Nucleon},
no additional ingredients are necessary.
Such study may provide an important theoretical 
input to the study of the reaction $p \bar p \to \pi^0 e^+ e^-$ in complement to
the first investigation presented here.

\vspace{0.3cm}
\noindent
{\bf Acknowledgments:}

\vspace{0.2cm}

The authors want to thank Beatrice Ramstein 
the comments and the careful reading of the manuscript. 
The authors thank Catarina Quintans and
Piotr  Salabura for helpful discussions.
G.~R.~was  supported by the Funda\c{c}\~ao para
a Ci\^encia e a Tecnologia under the Grant
No.~SFRH/BPD/26886/2006.
This work is also supported partially by the European Union
(HadronPhysics2 project ``Study of Strongly Interacting Matter'')
and by the  Funda\c{c}\~ao para a Ci\^encia e a Tecnologia,
under Grant No.~PTDC/FIS/113940/2009, 
``Hadron Structure with Relativistic Models''.

\appendix

\section{Kinematics}
\label{appA}

We consider here the final state of the decay process of a 
resonance $R$ according to
$R \to \gamma^\ast N$.
Assuming $W$ as the invariant mass of the resonance, 
we can write the four-momentum in the rest frame of $R$ as $P_R=(W,{\bf 0})$.
In this frame we can also write
\be
P_N=(E_N,-{\bf q}), \hspace{1.cm} 
q=(\omega, {\bf q}),
\ee
where $q$ is the photon momentum:
$P_R= P_N + q$, with $E_N=\sqrt{M^2+ {\bf q}^2}$ ($M$ is the nucleon mass).
We define then ${\bf q}$ as the photon three momentum
(symmetric to the nucleon three momentum) 
in the final state,
and $\omega$ as the photon energy in the $R$ rest frame.
All those variables are related with $W^2$ and $q^2$ 
according to
\ba
& & {\bf q}^2= \frac{(W^2+ M^2-q^2)^2}{4 W^2} - M^2 
\label{eqqvec2}
\\
& & \omega \equiv \frac{P_R \cdot q}{W}=
\frac{W^2- M^2 + q^2}{2W}.
\ea 
In the last relation we used
$E_N= \sfrac{W^2+ M^2-q^2}{2W}$.
From Eq.~(\ref{eqqvec2}) we can conclude that 
${\bf q}^2$ decreases when $q^2$ increases.
As ${\bf q}^2 \ge 0$, 
it can be proved that there is an upper limit to
$q^2$ (given by the condition ${\bf q^2}=0$).
The upper limit is then
\be
\left. q^2 \right|_{\rm max}= (W-M)^2.
\ee
This is then the largest value of $q^2$ 
for which the timelike form factors are defined.

In conclusion, the timelike form factors are 
defined only in a limited interval $[0,(W-M)^2]$.
One has then different ranges according  to the  value of $W$.
In the case $W=M$ we have only $q^2=0$.
At the pole $W =1.232$ GeV the maximum value of $q^2$
is 0.0859 GeV$^2$.
For $W=1.500$ GeV and $W=1.800$ GeV 
the maximum value of $q^2$ is respectively
0.315 GeV$^2$ and 0.741 GeV$^2$.

\end{document}